\documentclass[]{aa}
\usepackage{graphicx}
\usepackage{longtable, lscape}
\usepackage{natbib}
\begin{document}
\title{The globular cluster system of NGC1316. I. Wide-field photometry in the Washington system \thanks{Based
    on observations obtained  at the Cerro Tololo Inter-American Observatory, Chile}}

\subtitle{ }
\titlerunning{Globular cluster system of  NGC 1316}

\author{
T. Richtler     \inst{1} 
\and
L. P. Bassino \inst{2}
\and
 B. Dirsch \inst{3}
\and 
 B. Kumar \inst{4}
}

\institute{
Departamento de Astronom\'{\i}a,
Universidad de Concepci\'on,
Concepci\'on, Chile;
tom@astro-udec.cl
\and
Facultad de Ciencias
Astron\'omicas y Geof\'isicas de la Universidad Nacional de La Plata;
Consejo Nacional de Investigaciones Cient\'ificas y T\'ecnicas; and
Instituto de Astrof\'isica de La Plata (CCT La Plata-CONICET-UNLP),
Argentina 
\and
Friedrich-Ebert Gymnasium,
Bonn, Germany
\and
Arryabhatta Institute of Observational Science,
Nainital, India;
}

\offprints{T. Richtler}


\date{Received  / Accepted }

\abstract{NGC 1316 (Fornax A) is a prominent   merger remnant in the outskirts of the Fornax cluster. 
The bulge stellar population of NGC 1316  has a strong  intermediate-age component.
Studies of its globular cluster system may help to further refine its probably complex star formation history.
}
{ The  cluster system has not yet been  studied in its entirety.
We therefore present a wide-field study  of the globular cluster system of NGC 1316, investigating its properties in
relation to the global morphology of NGC 1316. }{We used the  MOSAIC II camera at the 4-m Blanco telescope at CTIO  in the filters Washington C and
Harris R. We identify globular cluster candidates and study their color distribution and the structural properties of the system. 
In an appendix, we also make morphological remarks, present color maps, and present new models for the brightness  and color profiles of the galaxy.}
{ The
cluster system is well confined to the optically visible outer contours of NGC 1316.  The total number of cluster candidates down to R=24 mag
is about 640. The color distribution of the entire sample is unimodal, but 
the color distribution of  bright subsamples in the bulge shows  two peaks that, by comparison with theoretical Washington colors with solar
metallicity, correspond to ages of about 2 Gyr and 0.8 Gyr, respectively.
We also find a significant population of 
clusters in the color range 0.8 $< $C-R $<$ 1.1 which must be populated by clusters younger than 0.8 Gyr, unless they are very metal-poor. 
The color interval 1.3 $<$ C-R $<$ 1.6 hosts the bulk of intermediate-age clusters which show a surface density profile 
with a sharp decline at about 4\arcmin.  The outer cluster population shows an unimodal color distribution with a peak
at C-R= 1.1, indicating a larger contribution of old, metal-poor clusters. However, their luminosity function does not show the expected turn-over, 
so the fraction of younger clusters is still significant.
We find a pronounced concentration of blue cluster candidates in the area of Schweizer's (1980) L1-structure. } 
{{Cluster formation in NGC 1316 has continued after an initial  burst, presumably related to the main merger. 
A toy model with two bursts of ages 2 Gyr and 0.8 Gyr is consistent with photometric properties and dynamical M/L-values. 
 In this model,
the older, metal-rich pre-merger population has an age of 7 Gyr,  contributes 90\% of the bulge mass and 70\% of the luminosity. Its
properties are  consistent  with
 spiral galaxies, where star-bursts were triggered by major/minor mergers and/or close encounters. }}


\maketitle

\section{Introduction}

The study of globular clusters systems (GCSs)  is motivated by the hope of finding clues to the formation
and history of their host galaxies (for reviews see \citealt{brodie06,harris10}). Globular cluster (GC) formation
occurs in a large variety of star forming environments and at all epochs.

 The formation of massive clusters is apparently favored in star bursts as they occur in
merger events (e.g. \citealt{whitmore95,degrijs03a}), but one finds GCs also in star-forming disks  of normal spiral galaxies where enhanced star
formation rates again seem to be related to an enhanced efficiency of massive cluster formation \citep{larsen99,larsen00}.
Against traditional wisdom,  intermediate-age  globular clusters exist even in the Milky Way \citep{davies11}. 
 
The old globular cluster systems of elliptical galaxies exhibit characteristic properties. A striking pattern is the color-bimodality, featuring
a blue (bona fide metal-poor) peak and a red (bona fide metal-rich) peak (e.g. \citealt{larsen01, kundu01}). 

Since early merger events are believed to be the driver for the formation of elliptical galaxies, the properties of GCSs of known
merger remnants may provide more insight  into the formation mechanisms.

The nearest merger remnant is NGC 5128 (Cen A)   at a distance of 3.8 Mpc.  Its stellar population is best described by
the bulk of stars (about 80\%) being very old, while a younger component (2-4 Gyr) contributes 20\%-30\% \citep{rejkuba11}. The GCS is  well investigated regarding kinematics, ages, and abundances \citep{peng04a,peng04b, woodley10a,woodley10b}.
Most GCs are old, but there are also intermediate-age and younger clusters. The kinematics of both planetary
nebulae and GCs indicate the existence of a massive dark halo \citep{woodley10a}.
Cen A is a ``double-double" radio galaxy (e.g. \citealt{saikia09}) with probably recurrent nuclear activity. 

Next in distance to Cen A is NGC 1316 (Fornax A)  in the outskirts of the Fornax cluster. There is a vast quantity of literature on NGC 1316 with studies in many 
wavelength bands. We  give a representative overview. 

 The  morphology  of NGC 1316 is very different from Cen A and is characterized by an
inner elliptical body with a lot of fine dust structure in its central region, best admired in HST images,  and an extended elongated 
structure with loops, tails, and tidal arms which cover almost an area on the sky comparable to the full moon. These were first described by \citet{schweizer80} in
a wide-field photographic study. \citet{schweizer81}  also noted the high central surface brightness. Furthermore, he  performed kinematical measurements and found an abnormally low $M/L_B = 1.8$ which in part is a consequence of his large adopted  distance of 32.7 Mpc.
More modern values are higher, e.g.  $M/L_V = 2.5$ \citep{shaya96} and $M/L_{K_s} = 0.65$ \citep{nowak08}, but still clearly indicate an intermediate-age
population.    \citet{schweizer80},  moreover, detected an inner ionizing rotating disk and a giant HII-region south of the center. He
 suggested a merger which occurred between 0.5 and 2 Gyr ago.

\citet{mackie98}, on the base of deep B-band imaging and imaging in H$_\alpha$ and NII, revisited the morphology and studied the
distribution of gaseous line emission. They detected faint emission to the North, between the nucleus and the companion galaxy
NGC 1317.  They also detected the interesting  ``Extended Emission Line Region (EELR)", an elongated feature of length 1.5\arcmin
South-West to the nucleus at a distance of about 6\arcmin, in a region without any signature of ongoing star formation.
Using ROSAT data, they detected hot gas apparently associated with Schweizer's tidal features L1 and L2 (see Fig.2 of 
\citealt{schweizer80}). Their general
assessment of NGC 1316's history is a disk-disk or disk-E merger older than 1 Gyr and a smaller merger about 0.5 Gyr ago.

\citet{horellou01} studied the content of atomic and molecular gas.  Molecules are abundant in the central region, but HI has been detected only in  some single spots, among them the EELR and the southern HII-region. 

NGC 1316 has been observed in the X-rays with practically all X-ray satellites (Einstein: \citealt{fabbiano92}; ROSAT:
\citealt{feigelson95}, \citealt{kim98};  ASCA: \citealt{iyomoto98}). More recently, \citet{kim03} used Chandra to constrain the
temperature of the hot ISM ($\approx$0.6 keV), to detect a low-luminosity AGN, and to detect 96 X-ray point sources. They
quote a range 0.25 Z$_\odot$ $<$ Z $<$ 1.3 Z$_\odot$ for
the metal-abundance of the ISM. The values derived from  Suzaku-data are lower than solar  \citep{konami10} (e.g. Fe is only 0.44 solar).
NGC 1316 also is among the sample of elliptical galaxies by \citet{fukuzawa06} who quote X-ray based dynamical masses. 
\citet{nagino09} used XMM-Newton to constrain the gravitational potential.

Kinematics of the stellar population and  abundances have been studied by \citet{kuntschner00}, \citet{thomas05}, and \citet{bedregal06}. The luminosity-weighted abundance is higher than
solar (Z $\approx$ 0.03).

\citet{arnaboldi98} presented kinematics of 43 planetary nebulae and long-slit spectroscopy along the major and the minor axis. They analyzed the velocity field
and gave some dynamical considerations, resulting in a total mass of $2.9~10^{11} M_\odot$ 
and a mass-to-light ratio in the B-band of 8 within 16 kpc. 

\citet{nowak08} investigated the very inner kinematics in order to constrain the mass of the supermassive black hole. They found
a mass of about  $1.5~10^8~M_\odot$ marginally consistent with the mass-sigma relation from \citet{tremaine02}.
Noteworthy is the double giant radio lobe with centers in projected galactocentric distances of   about 100 kpc (e.g. see Fig.1 of \citealt{horellou01}).
\citet{lanz10} used mid-infrared and X-ray data to develop a scenario in which the  lobes have been created by a nuclear outburst
about 0.5 Gyr ago. 

Very recently, \citet{mcneil12} presented radial velocities of almost 800 planetary nebulae in NGC 1316, one of the largest samples so far.  A spherical Jeans model indicates
a high dark matter content, characterized by a dark halo with a large core.

The GCS has been investigated several times with various intentions.  Apparently, the first observations devoted
to GCs were done using  HST (WF/PC-I) \citep{shaya96}.  These authors list 20 clusters in the very central region.
One object has $M_V = -12.7$ which, if old, today would be called an Ultracompact Dwarf (e.g. \citealt{mieske08}). 

More work on GCs, using the HST and the NTT, was presented by \citet{goudfrooij01a,goudfrooij01b,goudfrooij04}. Based
on infrared colors, they found ages for the brightest clusters consistent with 2-3 Gyrs. The luminosity function (LF)
of red GCs was found to be a power law with the exponent -1.2, flatter than that of normal ellipticals, while
the blue clusters exhibit the  normal LF with a turn-over at about $M_V \approx -7.2$ (but see our remarks in
Sec.\ref{sec:discussion_color}).

At the same time, \citet{gomez01} studied the GCS photometrically, using images obtained with the
3.6m telescope on La Silla, ESO, in B,V,I.   They did not detect a clear bimodality but a difference in the azimuthal 
distribution of blue and red clusters in the sense that the red clusters follow the ellipticity of the galaxy's bulge, while
the blue clusters are more circularly distributed. Moreover, they confirmed the very low specific frequency previously
found by \citet{grillmair99} of $S_N= 0.4$.

Our intention is to study the GCS on a larger field than  has been done before. Moreover,  our Washington photometry  permits a useful comparison
to the Washington photometry of elliptical galaxies, obtained with the same instrumentation
\citep{dirsch03a,dirsch03b,dirsch05,bassino06a}.
We adopt a distance of 17.8 Mpc , quoted by \citet{stritzinger10} using the four type Ia supernovae which appeared so far in NGC 1316.
The high supernova rate is a further indicator for an intermediate-age population.
The surface brightness fluctuation distance \citep{blakeslee09} is $21\pm0.6$ Mpc which may indicate a still unsolved zero-point problem.

This paper is the first in a series devoted to NGC 1316 and its globular cluster system. Future papers  
 will treat the kinematics
and dynamics of the GCS as well as 
 SH2, the HII-region
detected by \citet{schweizer80}.
 


\section{Observations and reductions}
\subsection{Data}
Since the present data have been taken and reduced together with the data leading to the
Washington photometry of NGC 1399 \citep{dirsch03b}, all relevant information regarding the
reduction technique can be found there. We thus give the most basic information only.

The  data
set consists of Washington wide-field images of NGC 1316 taken with
the MOSAIC camera mounted at the prime focus of the CTIO 4m Blanco telescope during the night 20/21 November 2001
 (the entire run had three nights). We used the Kron-Cousins R and Washington C filters. Although the genuine Washington
system uses T1 instead of R,  \citet{geisler96}  has shown that the Kron-Cousins R filter
is more efficient than T1, due to its larger bandwidth and higher throughput, and that R and
T1 magnitudes are closely related, with only a very small color term and a zero-point difference of 0.02 mag.

 The MOSAIC wide-field camera images a field of 36\arcmin $\times$ 36\arcmin, with a pixel scale of
0.27\arcsec /pixel. The observations were performed in the 16 channel read-out mode. We obtained three R exposures
with an exposure time of 600 sec each, and three C-exposures with an exposure time of 1200 sec each. Additionally,
we observed standard stars for the photometric calibration. In the following, we refer to colors as C-R and not as C-T1,
although the calibration provides Washington colors. The C-images have not been dithered.

The seeing was about 1\arcsec\  in the R-images and 1.5\arcsec\  in the C-images.

\subsection{MOSAIC reduction and photometry}
The MOSAIC data were reduced using the mscred package within IRAF. In particular this
software is able to correct for the variable pixel scale across the CCD which would cause otherwise a
4 percent variability of the brightness of stellar-like objects from the center to the corners. The flatfielding
resulted in images that had remaining sensitivity variations  of the order  of  1.5\%. 
In particular Chip  4 and
Chip 5 showed discernible remaining flatfield structure (but within the given deviation).

The actual photometry has been performed by using DAOPHOT II.

\subsection{Photometric calibration}
Standard fields for the photometric calibration have been observed in  each  
  of the 3 nights. The weather conditions were photometric. We observed 4-5 fields, each containing about 10 standard stars from
the list of \citet{geisler96}, with a large coverage of airmasses (typically from 1.0 to 1.9). It was
possible to use a single transformation for all three nights, since the coefficients derived for the
different nights were indistinguishable within the errors.
We derived the following relations between instrumental and standard magnitudes:

\begin{eqnarray} 
\nonumber
	\mathrm{R} =
	&\mathrm{R}_\mathrm{inst}+(0.72\pm0.01)-(0.08\pm0.01)X_\mathrm{R}\\
\nonumber
		&+(0.021\pm0.004)(\mathrm{C}-\mathrm{R})\\
\nonumber
	\mathrm{C} =	&\mathrm{C}_\mathrm{inst}+(0.06\pm0.02)-(0.30\pm0.01)X_\mathrm{C}\\
\nonumber
		&+(0.074\pm0.004)(\mathrm{C}-\mathrm{R})
\end{eqnarray}
The standard deviation of the difference between our calibrated and tabulated
magnitudes is 0.018\,mag in R and 0.027\,mag in C.

To calibrate the NGC\,1316 field we identified isolated stars which
were used to determine the zero points.  The scatter between the zero points determined from
individual stars is 0.03\,mag and most probably due to flat field
uncertainties.

The final uncertainties of the zero points are 0.03\,mag and 0.04\,mag for R
and C, respectively. This results in an absolute  calibration uncertainty in C-T1 of
0.05\,mag (the uncertainty in the color term can be neglected). See, however, Fig. \ref{fig:comparison}
which indicates that the calibration is quite precise in the interval 0.5 $< $C-R $<$ 2.5.

The foreground reddening towards NGC\,1316  according to 
\citet{schlegel98} is E$_{\mathrm{B-V}}=0.02$. Using
E$_{\mathrm{C-T1}}=1.97$\,E$_{\mathrm{B-V}}$ 
\citep{harris77} we had to correct C-R by 0.04\,mag. In the following we neglect the foreground reddening, since
no conclusion  depends strongly on an absolute precision of this order.

\subsection{Photometric uncertainties and selection of point sources}
\label{sec:selection}
\begin{figure}[h]
\begin{center}
\includegraphics[width=0.35\textwidth]{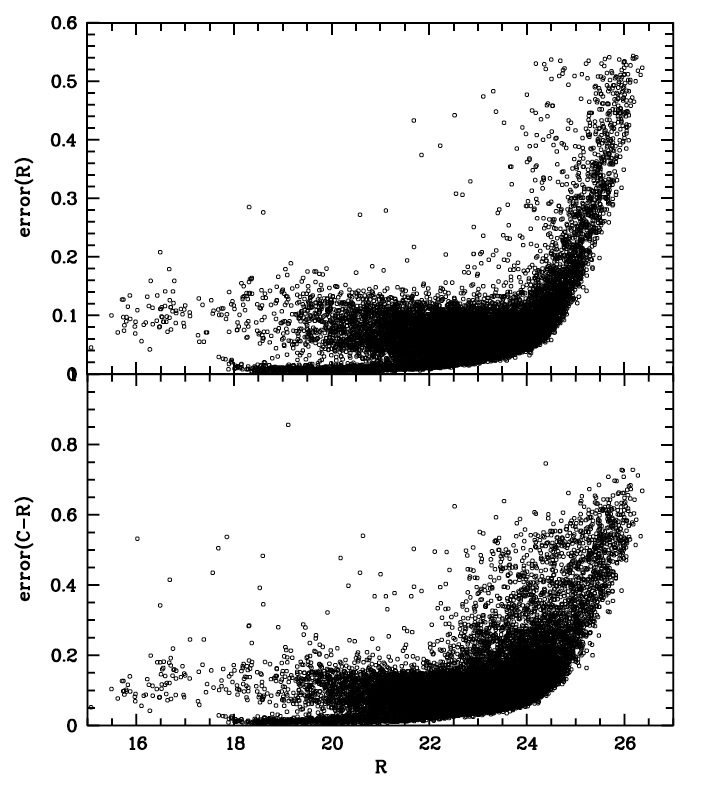}
\caption{Photometric uncertainties as given by DAOPHOT in R (upper panel) and C-R (lower panel). }
\label{fig:errors}
\end{center}
\end{figure}

Fig. \ref{fig:errors} shows the photometric errors as given by DAOPHOT for all sources. The upper panel
shows the uncertainties in R, the lower panel in C-R.
The point sources and resolved sources are clearly separated for magnitudes fainter than R=18.  Brighter point 
sources are saturated.  The uncertainties in color grow quickly for R$ >$ 24 mag. However, we want to select GC candidates not only on account of photometric uncertainties, but by goodness 
of fit and degree of resolution through the ALLSTAR parameters chi and sharp.

\begin{figure}[h!]
\begin{center}
\includegraphics[width=0.35\textwidth]{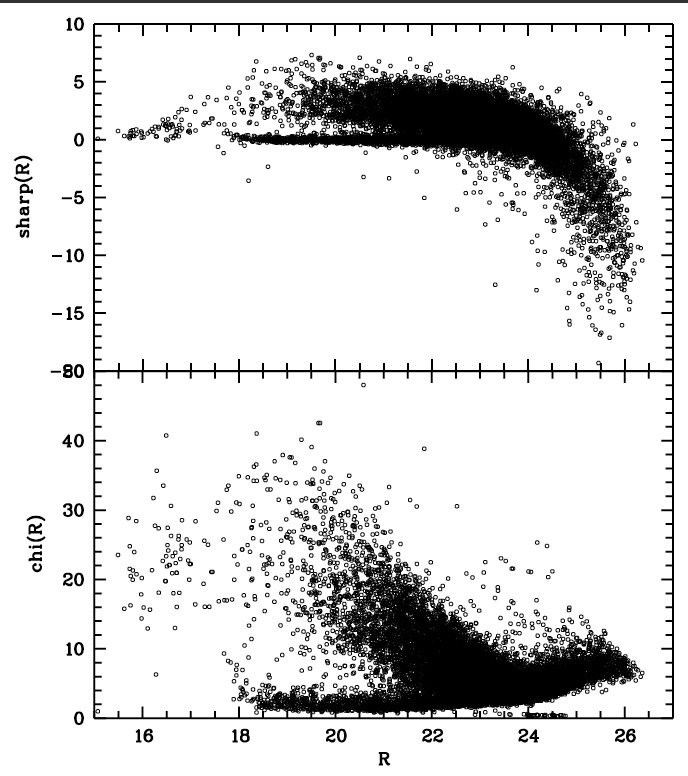}
\caption{The chi- and sharp parameters  of DAOPHOT.  We select as point sources objects within the sharp-parameter interval -0.5 $<$ sharp $<$ 0.8 and chi $<$ 4. }
\label{fig:chisharp}
\end{center}
\end{figure}

These parameters are plotted in Fig. \ref{fig:chisharp} with sharp in the upper panel and chi in the lower panel. Positive
values of sharp indicate resolved objects, negative values deficiencies in the photometry, mainly for the faintest objects.
The unsaturated resolved objects are of course galaxies, but we cannot exclude that some extended GCs are among
them (see the remark in Section \ref{sec:previousphot}).
 
\begin{figure}[h!]
\begin{center}
\includegraphics[width=0.35\textwidth]{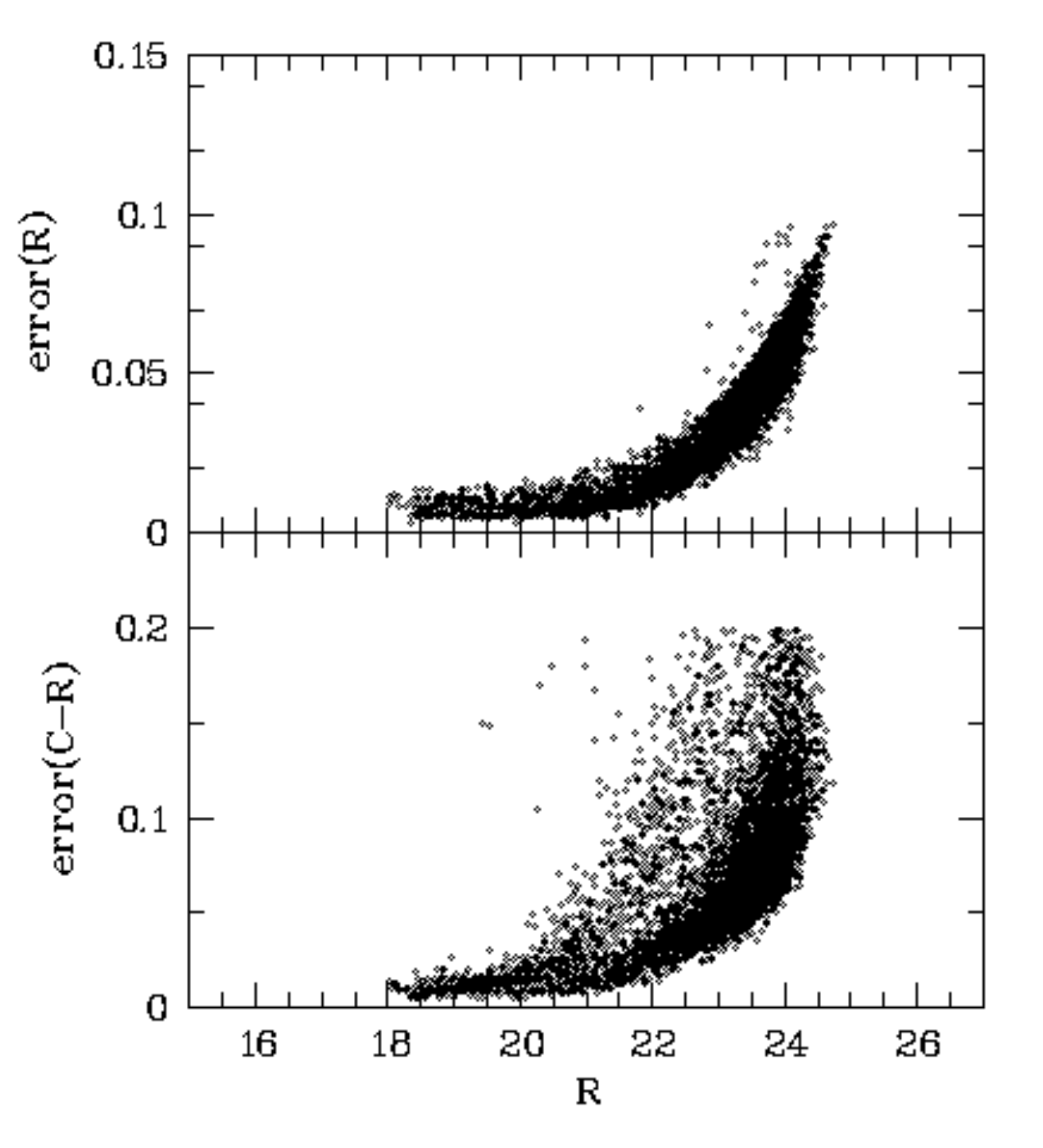}
\caption{Photometric uncertainties of selected point sources. Since the C-images are not dithered the larger errors at a given mag
are explainable by sources lying close to the CCD-gaps.}
\label{fig:resultingerrors}
\end{center}
\end{figure}

Guided by Fig. \ref{fig:chisharp}, we select as point sources objects within the sharp-parameter interval $-0.5 < sharp < 0.8$
and chi less than 4. 
We use the R-image due to its better seeing. Furthermore, we select the magnitude interval 18 $<$ R $<$ 24 and the
uncertainties in R and C-R to be less than 0.1 mag and 0.2 mag, respectively. This selection reduces the full sample
of about 22000 objects found in the entire field to 4675.
The resulting photometric errors for the so selected point sources are shown in Fig.\ref{fig:resultingerrors}.

\subsection{Comparison with previous photometry}
\label{sec:previousphot}
\begin{figure}[h]
\begin{center}
\includegraphics[width=0.35\textwidth]{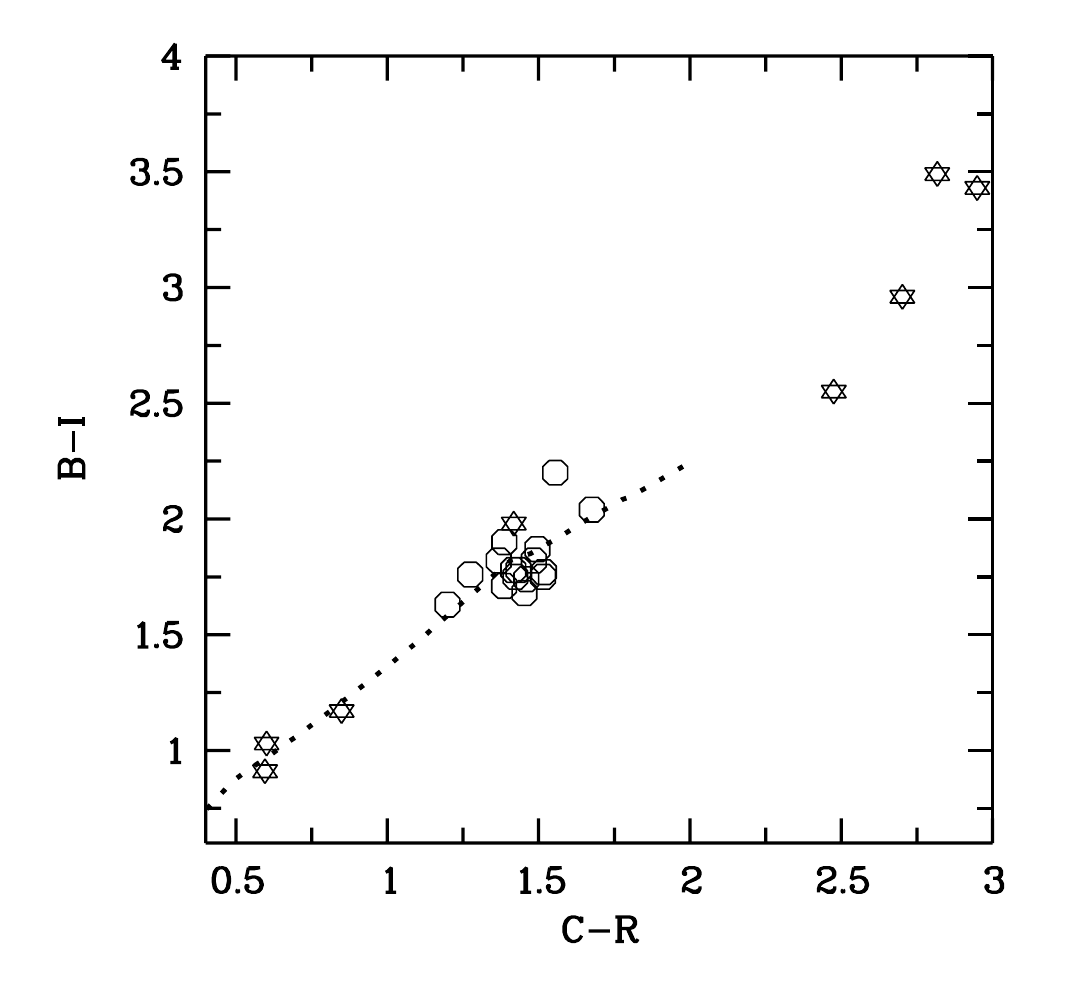}
\caption{Comparison of our C-R colors with the B-I colors of \citet{goudfrooij01a} for stars and confirmed globular clusters in NGC 1316. Star symbols denote stars,
circles denote clusters.  The dotted line delineates theoretical integrated single stellar population colors for solar metallicity, taken from \citet{marigo08}.}
\label{fig:comparison}
\end{center}
\end{figure}

We now compare for confirmed clusters in NGC 1316 our photometry with the B-I photometry of \citet{goudfrooij01a}. There are 17 clusters and 10 stars for which B-I and C-R colors
exist. Our point source selection excludes the object 211 of Goudfrooij et al. which seems to be very extended. 
Fig.\ref{fig:comparison}  shows the comparison. Star symbols denote stars, circles denote clusters.  For C-R colors bluer than 2, the agreement is very satisfactory with the exception of one cluster (the object 115 in Goudfrooij at al.'s list). This cluster is projected onto the immediate vicinity of a dust patch which may have influenced its
photometry.  The stars redder than C-R=2.5 deviate significantly which plausibly shows a deficiency of the photometric calibration which may be not valid for such red
objects. The dotted line represents the theoretical integrated single stellar population colors for solar metallicity, taken from \citet{marigo08}. Again, the agreement is very satisfactory.
  All clusters (except 115 of Goudfrooij et al. ) are located outside of the inner dust structures. However, only a strong reddening would
 be detectable.  The reddening vector in Fig.\ref{fig:comparison} is E(B-I) = 1.18~E(C-R), adopting the reddening law of \citet{harris77}, thus almost
 parallel to the (B-I)-(C-R)-relation.

\section{The  CMD of point sources}

\begin{figure*}[t]
\begin{center}
\includegraphics[width=0.3\textwidth,angle=-90]{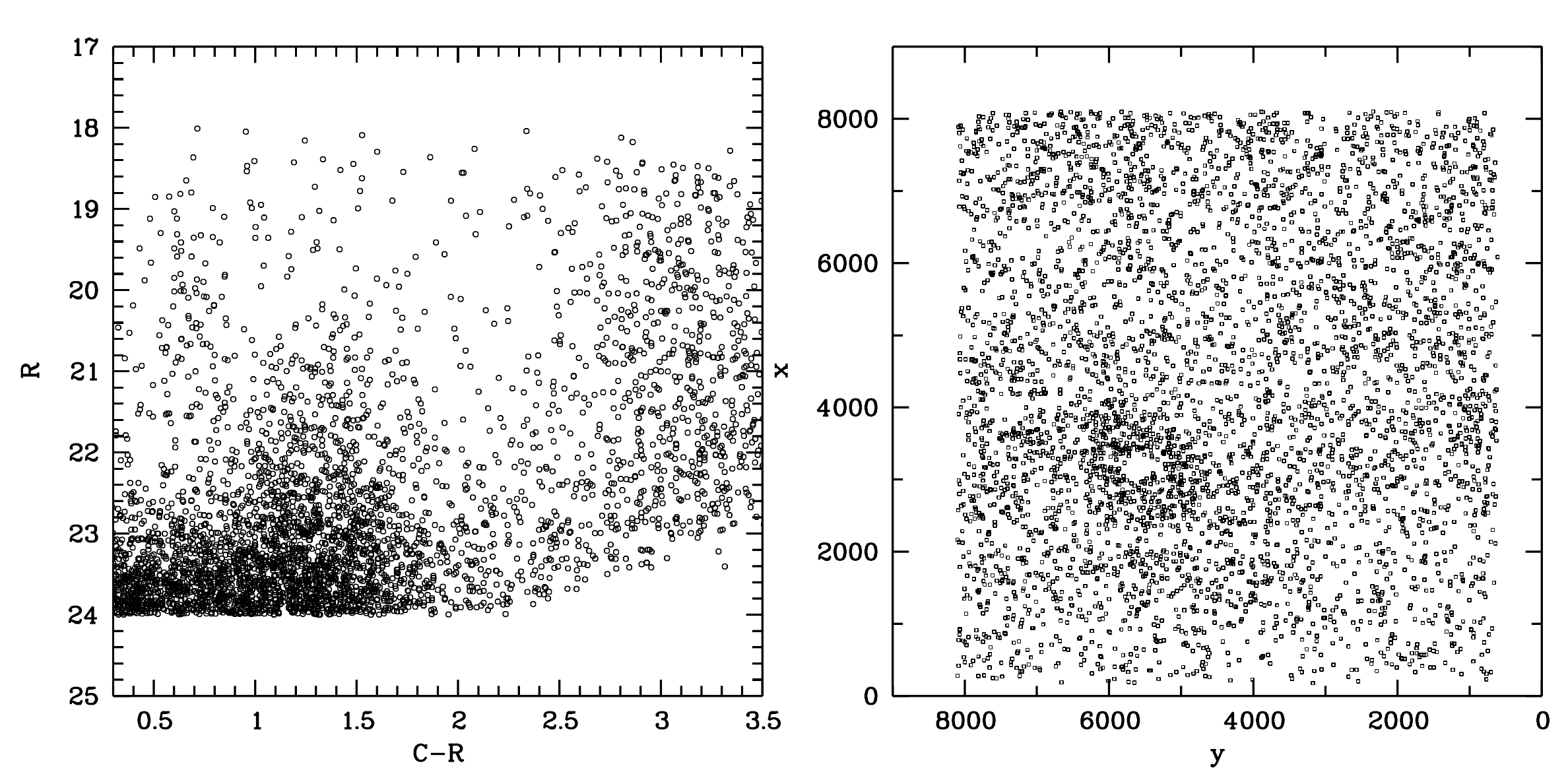}
\caption{Left panel: The CMD for 4675 point sources in the MOSAIC field around NGC 1316.   The bulk of the GCs are found in the color interval 1$<$C-R$<$2, but there are also bluer GCs. However,
most point sources bluer than C-R=1 and
23 $<$ R $<$ 24  are unresolved  background galaxies. Right panel: distribution of point sources in our MOSAIC field. The inner galaxy bulge is strikingly
visible. The size is 34\arcmin  $\times$ 34\arcmin. North is up, East to the left.}
\label{fig:CMDbig}
\end{center}
\end{figure*}

The CMD of the  selected 4675 point sources is shown in Fig.\ref{fig:CMDbig} (left panel) together with their location
(in pixels; right panel). Visible are the vertical stellar sequences at C-R=0.7 and C-R=3. The bulk of GC candidates
are found between C-R=1 and C-R=1.8 for R-magnitudes fainter than 20. There may be also GCs brighter than
R=20. The bulge of NGC 1316 is striking. We recall that the optically visible full diameter of NGC 1316 is about 27\arcmin, corresponding
to 6000 pixels. At first glance there is no bimodality.  A more detailed look into color intervals will modify this impression.
R=18 mag is approximately the saturation limit for point sources. A comparison with the field around NGC 1399 \citep{dirsch03b}
(their Fig.3) shows that the present data are considerably deeper. At a magnitude level of R=24 mag, the incompleteness
due to faint sources in C starts at  C-R=1.3, while around NGC 1316, it starts at C-R=2.3. 

The inner blank region of NGC 1316 shows the incompleteness where the galaxy light gets too bright. To avoid this
incompleteness region, we shall consider only GC candidates with distances larger than 2\arcmin.

\begin{figure*}[t]
\begin{center}
\includegraphics[width=0.8\textwidth,angle=0]{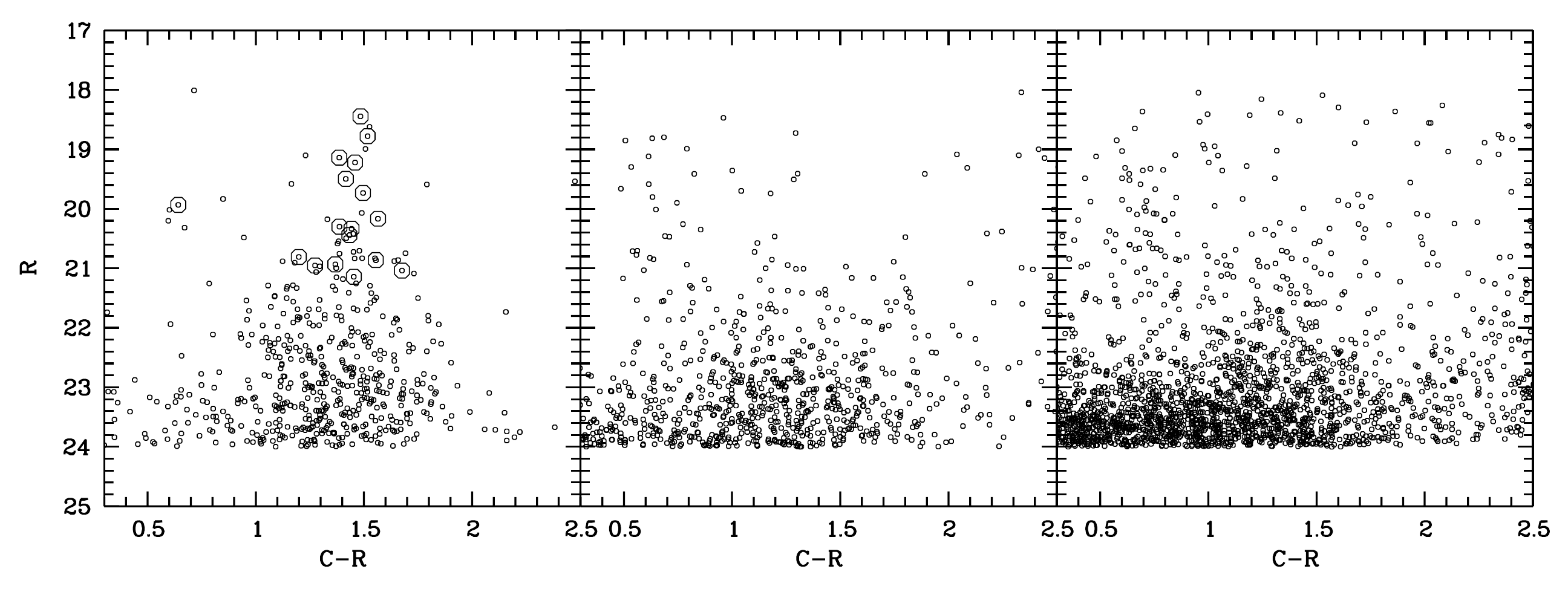}
\caption{The three CMDs refer to different radius selections. The left panel shows GC candidates within the inner 4\arcmin, the middle
panel between 4\arcmin\ and 10\arcmin, and the outer panel at radii larger than 10\arcmin.  The respective area sizes then  are related
as 1:5.2:25.3.  The inner CMD is thus dominated by genuine GCs. In the left panel, spectroscopically confirmed GCs \citep{goudfrooij01b}
are marked. Note the blue object at C-R = 0.64 which must be a quite young cluster.}
\label{fig:CMD_radius_selected}
\end{center}
\end{figure*}

Fig. \ref{fig:CMD_radius_selected}  shows  CMDs  with three different radius selections (see figure caption). In the inner  region 
(which in fact is blank within the inner 30\arcsec; here we consider also clusters closer to the center than 2\arcmin) the vertical sequence of bright objects at C-R=1.5 is striking. Some
of these GC candidates are already confirmed clusters from the radial velocity sample of \citet{goudfrooij01b}. 
Very interesting is the blue object with C-R=0.64 (object 119 in Goudfrooij et al.'s list) which must be quite young, about 0.5 Gyr (see Sec. \ref{sec:discussion_color}),
assuming solar metallicity.
 Two more objects with very red colors, which appear 
as GCs in the list of Goudfrooij et al. (numbers 122 and 211), are not  marked. With R=20.37, C-R = 3.26, and R= 20.77, C-R=3.57,
they fall drastically outside the color range of GCs. They are not obviously associated with dust.  On our FORS2/VLT pre-images (seeing about 0.5\arcsec), used for 
the spectroscopic campaign, they are
clearly resolved and might be background galaxies, in which  case the velocities are incorrect.

However,  if one  bright young cluster exists, then we expect also fainter ones of similar age. We shall show that statistically.
The middle panel demonstrates that the occurrence of these bright clusters is restricted to the inner region, but there are still
overdensities of GC candidates. The right panel shows almost exclusively  foreground stars and background galaxies.  
 We now have a closer look at the distribution of colors.

\section{Color distribution}
\label{sec:color}
\begin{figure*}[]
\begin{center}
\includegraphics[width=0.6\textwidth]{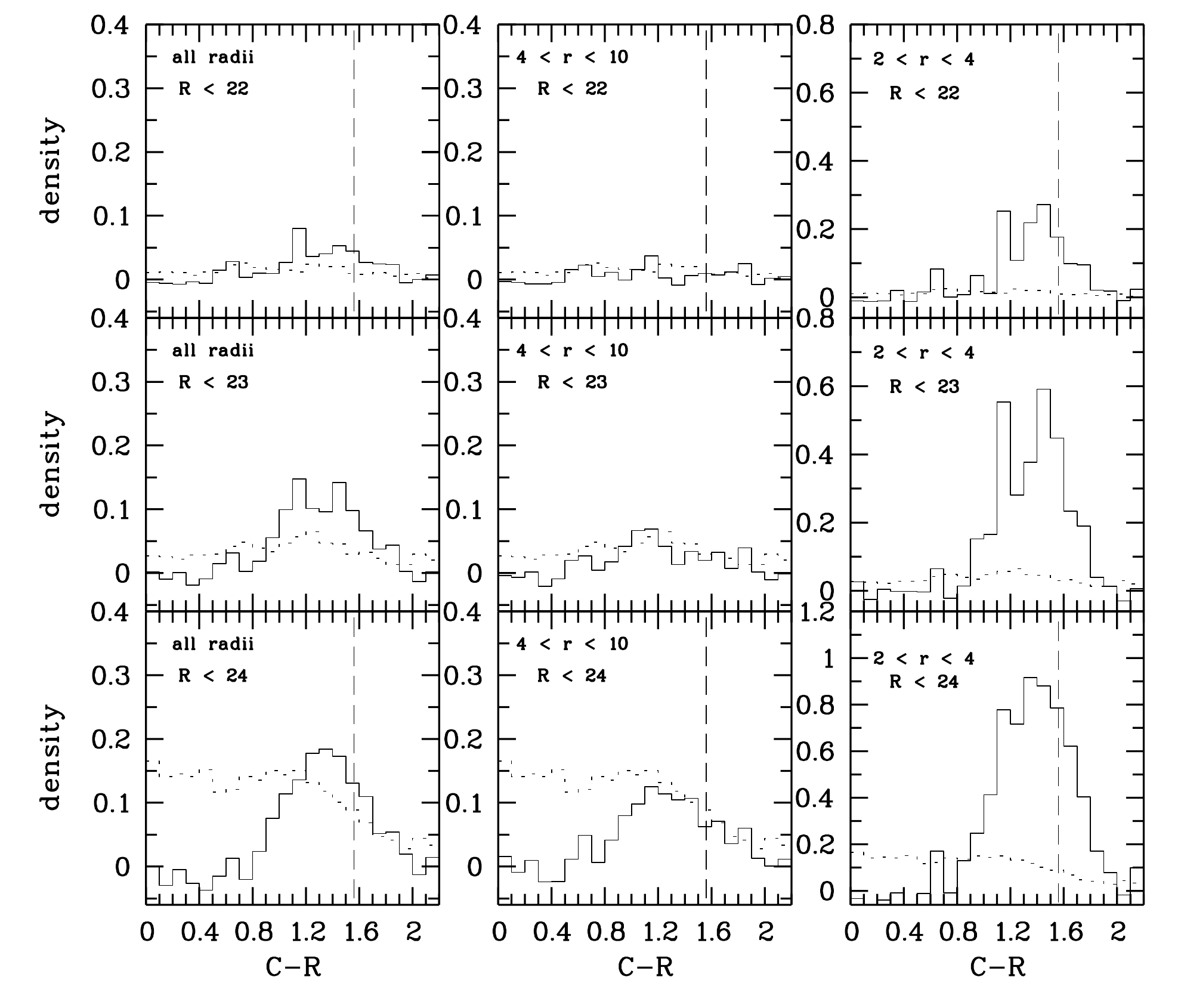}
\caption{Color histograms, corrected for background counts, for three radial regimes and for different magnitude ranges. Left panel: 0-10 arcmin, middle panel:
4-10 arcmin, right panel: 2-4 arcmin. The upper row displays magnitudes brighter than R=22, the middle row objects brighter than R=23, the lower row
objects brighter than R=24. The respective backgrounds are indicated by the dotted histograms. 
The vertical dashed line indicates the color of NGC 1316 at 2\arcmin, outside the dusty region. 
}
\label{fig:colordistribution}
\end{center}

\end{figure*}

Fig. \ref{fig:colordistribution} displays the color distribution in different radial regimes (left: radius $<$ 10\arcmin; middle: 4\arcmin $<$ radius $<$ 10\arcmin; right: 2 $<$ radius $<$ 4\arcmin)  for all magnitudes  (R $<$ 24 mag)(lower row) , for bright magnitudes (R $<$ 22) (upper row),
and for an intermediate sample magnitudes (R $<$ 23) (middle row). A radius of 4\arcmin\ approximately marks the extension of the bulge. The ordinates are given in numbers per arcmin$^2$.
The corresponding values and their uncertainties are given in Table \ref{tab:color24} for the case   R $<$ 24 mag and in Table \ref{tab:color23}
for the case R$<$ 23 mag.

 The counts are background corrected, the background being evaluated outside 13\arcmin.

The main observations from Fig.\ref{fig:colordistribution} are:\\
\noindent
-- the entire system shows a continuous distribution of colors with no sign of bimodality\\
-- the peak of the color distribution is displaced bluewards with respect to the galaxy color,  more so for the outer clusters\\
-- the inner bright clusters  show a pronounced bimodality with peaks  at C-R=1.4 and C-R=1.1. These peaks are already indicated
   in the small, but clean, spectroscopic sample of \citet{goudfrooij01a}\\
-- the outer clusters show only a peak a C-R=1.1\\
-- there is evidence for a small GC population bluer than C-R=1.0 which is approximately the metal-poor limit for old clusters (see Fig.\ref{fig:washington}).\\

The color distribution of GCs in a typical elliptical galaxy in our photometric system is bimodal with two peaks at  C-R=1.35
and C-R=1.75 \citep{bassino06b} with little scatter. A Gauss fit to the entire sample in the
color range 0.8-2 gives  a peak color
of $1.33 \pm 0.01$ and a sigma of $0.3\pm 0.01$ which resembles indeed the position of the ``universal'' blue peak color in the GCS of elliptical galaxies 
while the distribution is broader. 
On first sight, this is somewhat surprising since we expect
a  quite different composition of GCs: there should be a  mixture of old GCs from the pre-merger components, and an unknown fraction of GCs, presumably a large one,
 which stem from one or more star-bursts. Later star forming-events with GC production might also have occurred. 
 The facts that the already confirmed clusters define such a sharp color peak (Fig. \ref{fig:CMD_radius_selected}) and that the existence of intermediate-age clusters has already
 been shown, strongly suggest that among the bright clusters, intermediate-age objects are
 dominating. Older clusters are then cumulatively mixed in with decreasing brightness.  
 
   The peak color for the outer sample is $1.23\pm0.037$ and for the inner sample $1.37\pm0.01$.
This difference must, in part, be due to a different composition of GC populations since the density of red clusters shows a steeper decline, but we
cannot exclude a contribution from differential reddening.   However, this must be small since there are hardly any cluster candidates
redder than C-R=1.9. Moreover, the excellent reproduction of the theoretical color-color relations in Fig. \ref{fig:comparison}
indicates the absence of {\bf strong} reddening for this inner sample where we would expect to see reddening effects first.
 
\begin{table}[h!]
\caption{Background subtracted numbers per arcmin$^2$ of GC candidates brighter than R=24 mag in color bins and for different radial selections.Bin widths are 0.1 mag. The background is defined by r$>$13\arcmin. The uncertainties are based on the
square root of the raw counts. Negative values are kept for formal correctness.}
\begin{center}
\resizebox{9cm}{!}{
\begin{tabular}{ccccc}
Color &  background & 2$<$r$<$10 & 4$<$r$<$10 & 2$<$r$<$4\\
\hline
   0.650 & 0.144 $\pm$ 0.014 & 0.041 $\pm$ 0.028 & 0.026 $\pm$ 0.029 & 0.148 $\pm$ 0.089\\
 0.750 & 0.167 $\pm$ 0.015 & -0.01 $\pm$ 0.026 & -0.01 $\pm$ 0.028 & -0.03 $\pm$ 0.061\\
 0.850 & 0.155 $\pm$ 0.014 & 0.033 $\pm$ 0.028 & 0.023 $\pm$ 0.030 & 0.111 $\pm$ 0.085\\
 0.950 & 0.167 $\pm$ 0.015 & 0.085 $\pm$ 0.032 & 0.064 $\pm$ 0.033 & 0.231 $\pm$ 0.104\\
 1.050 & 0.159 $\pm$ 0.015 & 0.141 $\pm$ 0.034 & 0.083 $\pm$ 0.034 & 0.398 $\pm$ 0.122\\
 1.150 & 0.163 $\pm$ 0.015 & 0.216 $\pm$ 0.038 & 0.113 $\pm$ 0.035 & 0.766 $\pm$ 0.158\\
 1.250 & 0.156 $\pm$ 0.014 & 0.214 $\pm$ 0.037 & 0.089 $\pm$ 0.034 & 0.693 $\pm$ 0.151\\
 1.350 & 0.138 $\pm$ 0.014 & 0.225 $\pm$ 0.037 & 0.084 $\pm$ 0.032 & 0.896 $\pm$ 0.166\\
 1.450 & 0.121 $\pm$ 0.013 & 0.226 $\pm$ 0.036 & 0.087 $\pm$ 0.031 & 0.860 $\pm$ 0.162\\
 1.550 & 0.098 $\pm$ 0.011 & 0.176 $\pm$ 0.032 & 0.053 $\pm$ 0.026 & 0.777 $\pm$ 0.153\\
 1.650 & 0.077 $\pm$ 0.010 & 0.137 $\pm$ 0.028 & 0.063 $\pm$ 0.025 & 0.613 $\pm$ 0.136\\
 1.750 & 0.058 $\pm$ 0.009 & 0.076 $\pm$ 0.022 & 0.025 $\pm$ 0.020 & 0.393 $\pm$ 0.110\\
 1.850 & 0.047 $\pm$ 0.008 & 0.077 $\pm$ 0.021 & 0.055 $\pm$ 0.021 & 0.165 $\pm$ 0.075\\
 1.950 & 0.036 $\pm$ 0.007 & 0.015 $\pm$ 0.015 & 0.005 $\pm$ 0.014 & 0.070 $\pm$ 0.054\\
 2.050 & 0.054 $\pm$ 0.009 & -0.01 $\pm$ 0.014 & -0.00 $\pm$ 0.016 & -0.02 $\pm$ 0.028\\
 2.150 & 0.036 $\pm$ 0.007 & 0.021 $\pm$ 0.015 & 0.009 $\pm$ 0.015 & 0.096 $\pm$ 0.060\\
 \hline
\end{tabular}
}
\end{center}
\label{tab:color24}
\end{table}%

\begin{table}[h!]
\caption{Background subtracted numbers per arcmin$^2$ of GC candidates brighter than R=23 mag in color bins and for different radial selections. Bin widths are 0.1 mag. The background is defined by r$>$13\arcmin. The uncertainties are based on the
square root of the raw counts. Negative values are kept for formal correctness.}
\begin{center}
\resizebox{9cm}{!}{
\begin{tabular}{ccccc}
Color &  background & 2$<$r$<$10 & 4$<$r$<$10 & 2$<$r$<$4\\
\hline 
0.650 & 0.042 $\pm$ 0.007 & 0.032 $\pm$ 0.017 & 0.026 $\pm$ 0.018 & 0.064 $\pm$ 0.054\\
 0.750 & 0.048 $\pm$ 0.008 & 0.003 $\pm$ 0.015 & 0.004 $\pm$ 0.016 & -0.02 $\pm$ 0.028\\
 0.850 & 0.039 $\pm$ 0.007 & 0.018 $\pm$ 0.015 & 0.018 $\pm$ 0.016 & 0.014 $\pm$ 0.038\\
 0.950 & 0.034 $\pm$ 0.007 & 0.056 $\pm$ 0.018 & 0.042 $\pm$ 0.018 & 0.152 $\pm$ 0.071\\
 1.050 & 0.047 $\pm$ 0.008 & 0.099 $\pm$ 0.023 & 0.066 $\pm$ 0.022 & 0.165 $\pm$ 0.075\\
 1.150 & 0.056 $\pm$ 0.009 & 0.147 $\pm$ 0.027 & 0.068 $\pm$ 0.023 & 0.554 $\pm$ 0.128\\
 1.250 & 0.065 $\pm$ 0.009 & 0.101 $\pm$ 0.025 & 0.041 $\pm$ 0.022 & 0.280 $\pm$ 0.096\\
 1.350 & 0.047 $\pm$ 0.008 & 0.096 $\pm$ 0.023 & 0.013 $\pm$ 0.017 & 0.377 $\pm$ 0.106\\
 1.450 & 0.046 $\pm$ 0.008 & 0.142 $\pm$ 0.026 & 0.034 $\pm$ 0.019 & 0.591 $\pm$ 0.130\\
 1.550 & 0.030 $\pm$ 0.006 & 0.098 $\pm$ 0.021 & 0.020 $\pm$ 0.015 & 0.448 $\pm$ 0.113\\
 1.650 & 0.032 $\pm$ 0.007 & 0.066 $\pm$ 0.019 & 0.032 $\pm$ 0.017 & 0.233 $\pm$ 0.084\\
 1.750 & 0.023 $\pm$ 0.006 & 0.038 $\pm$ 0.015 & 0.007 $\pm$ 0.012 & 0.189 $\pm$ 0.075\\
 1.850 & 0.013 $\pm$ 0.004 & 0.044 $\pm$ 0.014 & 0.039 $\pm$ 0.015 & 0.040 $\pm$ 0.038\\
 1.950 & 0.013 $\pm$ 0.004 & 0.002 $\pm$ 0.008 & 0.002 $\pm$ 0.009 & 0.013 $\pm$ 0.027\\
 2.050 & 0.030 $\pm$ 0.006 & -0.01 $\pm$ 0.010 & -0.01 $\pm$ 0.011 & -0.03 $\pm$ 0.006\\
 2.150 & 0.020 $\pm$ 0.005 & 0.002 $\pm$ 0.010 & -0.00 $\pm$ 0.010 & 0.006 $\pm$ 0.027\\
 \hline
\end{tabular}
}
\end{center}
\label{tab:color23}
\end{table}%

\section{Spatial distribution of clusters}
\subsection{Density profile and total numbers}
\begin{figure}[h!]
\begin{center}
\includegraphics[width=0.4\textwidth]{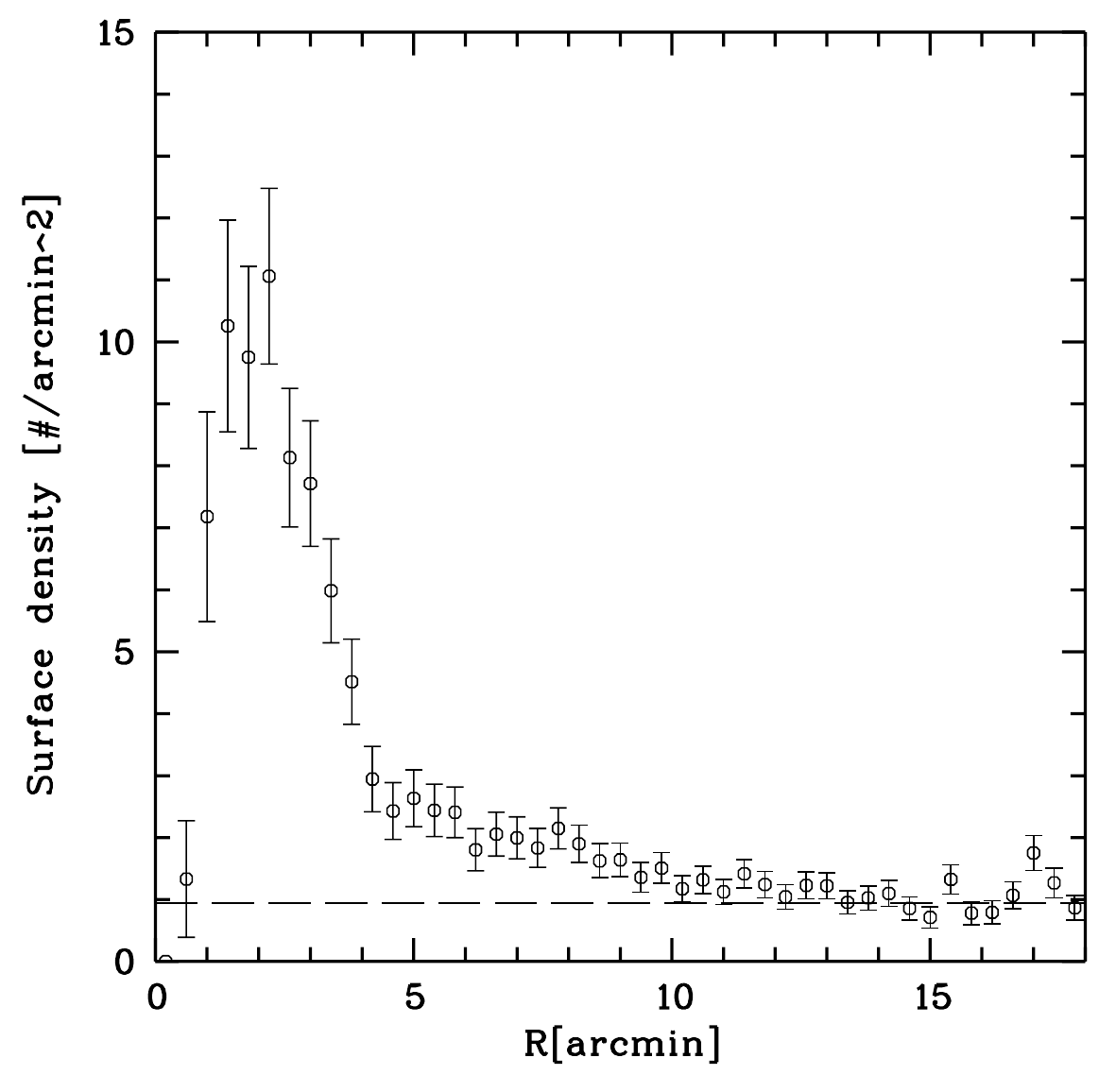}
\caption{Surface densities for selected point sources. The dashed horizontal line indicates the background. Inside  2\arcmin, the counts are severely affected by incompleteness. The profile
seems to be inflected at a radius of 4\arcmin.  We define the background by counts outside 13\arcmin.}
\label{fig:profile_all}
\end{center}
\end{figure}

Given the complicated morphology and a probable mixture of bulge and disk symmetry, a surface density profile for GCs in NGC 1316 will not have the same meaning
as in spherical elliptical galaxies, where deprojection is possible. We rather use the density profile to
evaluate the total extent of the cluster system and possible features in the profile. For convenience we
count in circular annuli and select  point sources in the color range 0.9-2 according to the above selection.
The full annuli become sectors for distances larger than 10\arcmin, when the image border is reached. We correct for this geometrical incompleteness.

First we show the general density profile, then the two-dimensional distribution in various color intervals, and finally the
radial density profiles in these intervals.

Fig. \ref{fig:profile_all} shows the resulting surface densities. Inside 2\arcmin\ the counts are severely affected by incompleteness
due to the galaxy light and possible extinction by dust. Outside 13\arcmin\  the counts are  not distinguishable anymore from the background, for which we
 determine 0.946 objects/arcmin$^2$.
 
Summing up the bins in Fig.\ref{fig:profile_all}, we find  636 $\pm$ 35 as the total number of GC candidates down to our magnitude limit of R=24 and in the color range 0.9 $<$ C-R $<$ 2.0 inside 13\arcmin. Correcting for the
 inner incompleteness would not enhance this number considerably. In case of  an old GCS of a normal elliptical galaxy with a
 Gaussian-like luminosity function  and a turn-over magnitude (TOM) corresponding to its distance, one would roughly double this number to have a fair estimation of the total number.
 Since a  single TOM, valid for the entire cluster system, does not exist, one would underestimate 
 this number. About 1400-1500 clusters seems to be a good guess, somewhat larger than the number given by \citet{gomez01}.
 In any case, it is a relatively poor cluster system. We come back to this point in the discussion. 



This radius of 13\arcmin\ agrees well with the extension of NGC 1316 seen in Fig.\ref{fig:picture} which we expect if the extension of
NGC 1316 is determined by the dynamical processes during the merger event(s). One notes an inflection point in Fig.\ref{fig:profile_all}
at about  4\arcmin. Its nature becomes clearer if we subtract the background and plot surface densities for different
color intervals. This is done in Fig.\ref{fig:profile_density_linear} in a linear display, including the interval 0.8$<$C-R$<$1.1,
and in Fig.\ref{fig:profile_colors} in double-logarithmic display.  The existence of  a blue population is statistically well visible, although
poor. GC candidates in the color interval 1.3-1.6 show a sharp decline at 4\arcmin, which defines the bulge. The
outer clusters in this sample are mainly located to the South.

\subsection{The two-dimensional distribution of clusters}
\label{sec:2dim}
\begin{figure*}[ht]
\begin{center}
\includegraphics[angle=-90,width=0.7\textwidth]{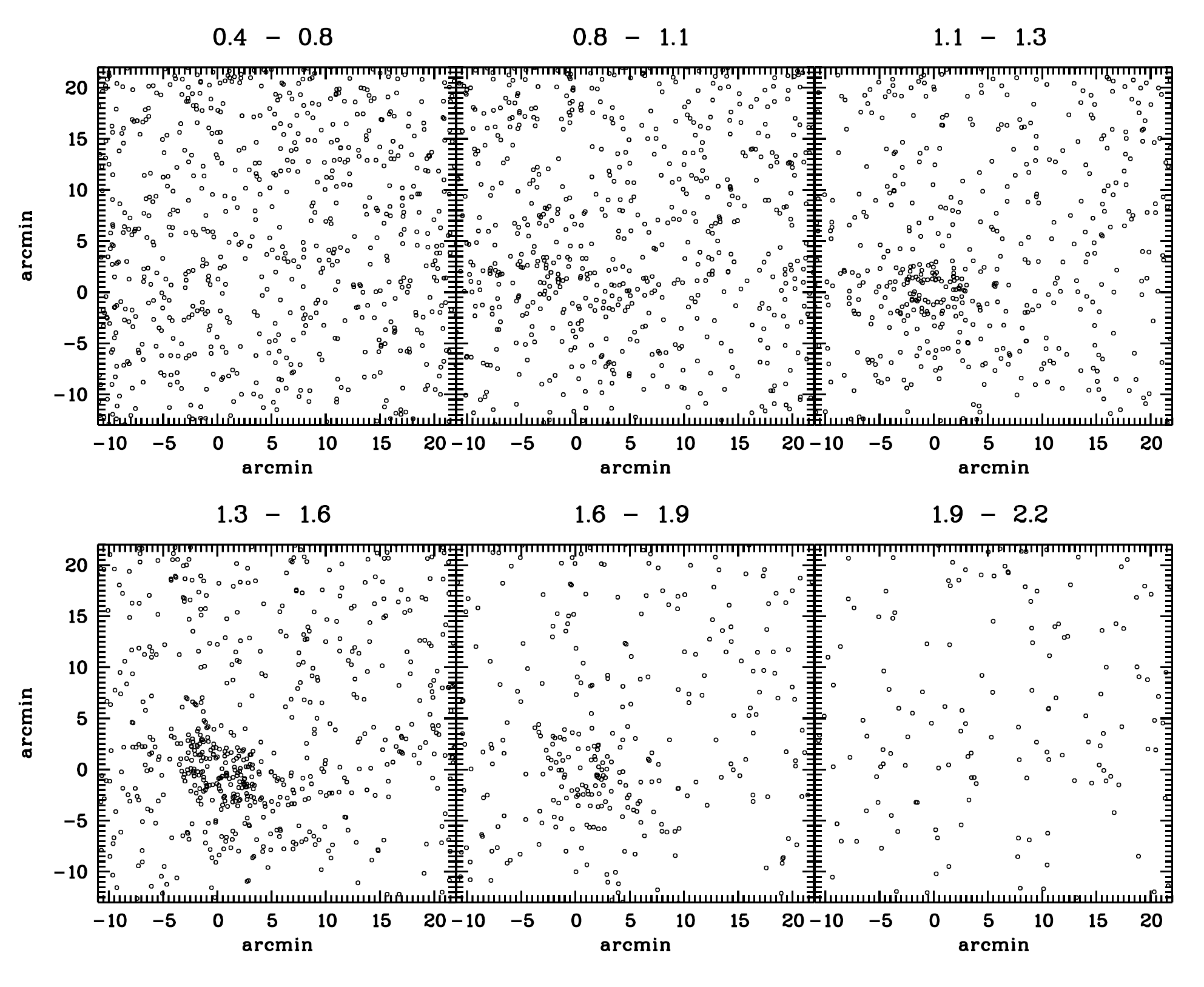}
\caption{This plot shows the 2-dimensional distribution of GC candidates in six  C-R color intervals. North is up, East to the left. The origin
is NGC 1316. See text for details. The most
important observations are: there are clusters bluer than C-R=1.1, clusters in the range 1.1-1.3 show a symmetry with a shifted major axis,
clusters of intermediate color define mostly the bulge but there is also an overdensity of clusters related to outer morphological features to the
South.  }
\label{fig:xyplot_color}
\end{center}
\end{figure*}

NGC 1316 exhibits an elliptical symmetry only in its inner parts. At larger radii, 
the overall morphology is characterized by tidal(?) tails and loops.
It is now interesting to
ask whether morphological features of NGC 1316 can be found back in the distribution of GCs. Naturally, one does that
in color intervals to suppress the background and  (hopefully) get insight into the nature of the cluster candidates.

We plausibly assume that intermediate-age and younger clusters have at least the metallicity of the stellar bulge, i.e. at
least solar metallicity. Refer to Fig.\ref{fig:washington} for a theoretical relation between age and Washington colors. 

Fig.\ref{fig:xyplot_color} shows the 2-dimesional distribution of objects in six different color-intervals.

Interval 0.4$<$ C-R $<$0.8:\\
Here we expect younger clusters with ages about 0.5 Gyr or somewhat younger. NGC 1316 is statistically not visible but  note the cluster with C-R = 0.6 in Fig.\ref{fig:CMD_radius_selected}. We 
anticipate (Richtler et al., in prep.) that we identified another  object with C-R=0.4 by its radial velocity.  Certainly there are not
many bright clusters within this age range and they do not seem to belong to a particularly strong star formation epoch. 

Interval 0.8$<$ C-R $<$1.1\\
Now the field of NGC 1316  becomes  recognizable. With solar metallicity or higher,  ages are between 0.6 and 1 Gyr. Noteworthy are the clumps roughly 1.5\arcmin\ to the NE
and a slight overdensity which is projected roughly onto the L1-feature. 

Interval 1.1$<$ C-R $<$1.3:\\
In this interval we expect old, metal-poor clusters and  younger clusters between 1 and 1.5 Gyr. NGC 1316 is now clearly
visible. The major axis of the distribution  seems to show a position angle of about 90$^\circ$, different from the main
body. However, the more quantitative source counts in dependence on azimuth still show an overdensity at the
position angle of the major axis of NGC 1316 towards the North-East, while the peak towards the South-West is less
pronounced (Fig. \ref{fig:azimut}). 
The fraction of intermediate-age clusters is unknown. We shall discuss this issue further in Sec. \ref{sec:subpop}, where we
argue that  younger clusters actually  provide the dominating population.

Interval 1.3$<$ C-R $<$1.6\\
This interval samples the maximum of the color distribution.  One notes the sharp definition of the galaxy's luminous body. Perpendicular to
the major axis the cluster reaches the background  at about  2.5\arcmin. A gap is visible at a distance of approximately
4\arcmin\ to the South. Even further to the South, the cluster candidates populate a region roughly  defined 
by Schweizer's L1-structure. 
  Note also  the 
horizontal boundary  of cluster candidates at a distance of 8\arcmin\ which delineates nicely the sharp ridge in NGC 1316. 

Interval 1.6$<$ C-R $<$1.9\\ 
Here we expect the majority of the metal-rich GCs of the pre-merger components. The fact that NGC 1316 is still well visible,
indicates a similar concentration as that of the metal-poor clusters. The overall symmetry follows the main body of NGC 1316,
which indicates the presence of  an old population. In fact, as we argue in Section \ref{sec:scenario}, this population dominates
the bulge mass.

Interval 1.9$<$ C-R $<$2.2\\ 
One would expect old, metal-rich, and reddened clusters in this interval. However,  NGC 1316 is not longer visible.  

\subsection{Density profiles in various color intervals}

\begin{figure}[h!]
\begin{center}
\includegraphics[width=0.4\textwidth]{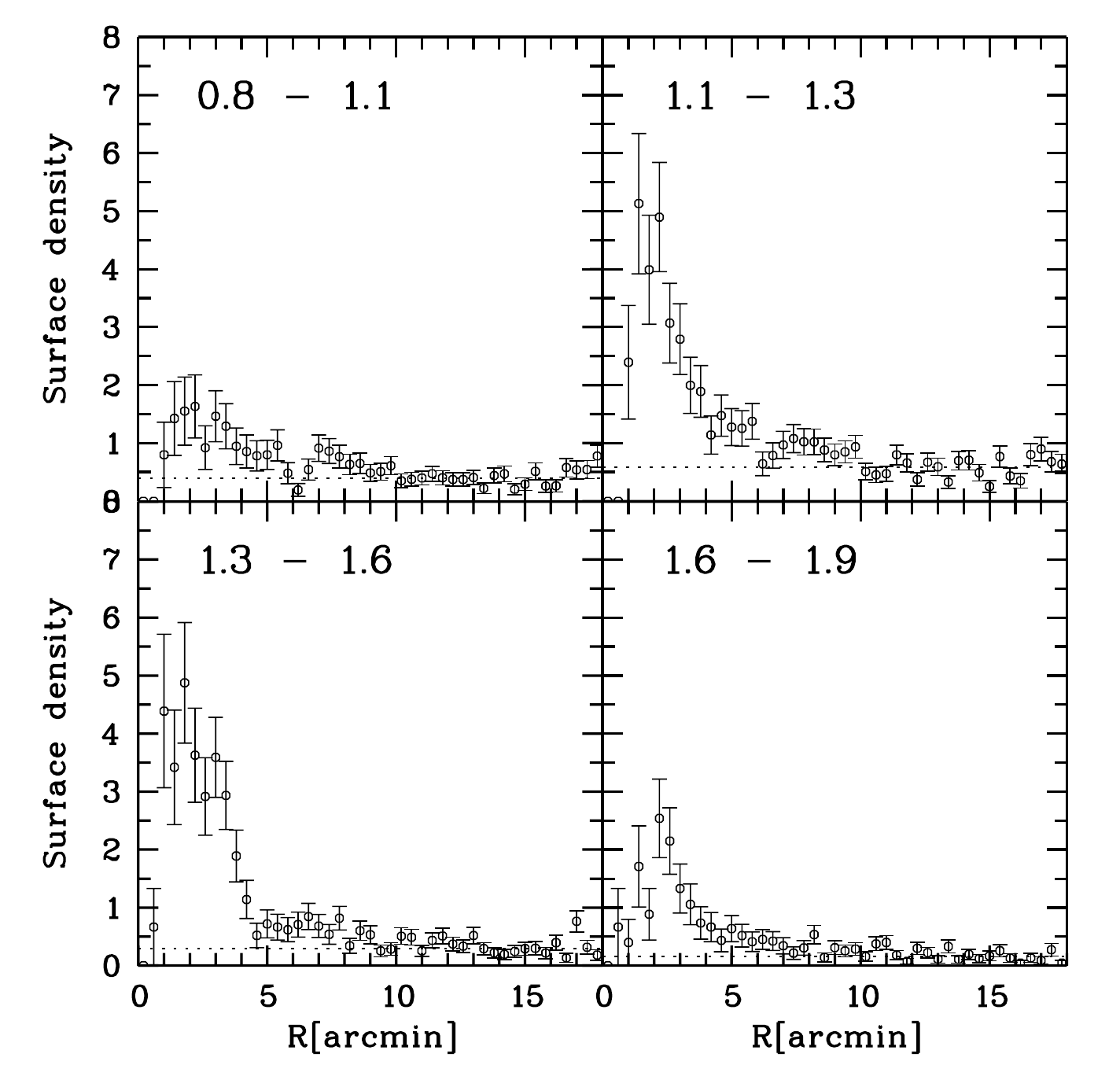}
\caption{Background subtracted surface densities for selected point sources in different C-R color intervals and linearly displayed. 
See Sec.\ref{sec:2dim} for the significance of these intervals and Fig.\ref{fig:profile_colors} for a logarithmic display. Although
only a poor population, clusters bluer than 1.1 are statistically well visible.  }
\label{fig:profile_density_linear}
\end{center}
\end{figure}

\begin{figure*}[t]
\begin{center}
\includegraphics[width=0.8\textwidth,angle=0]{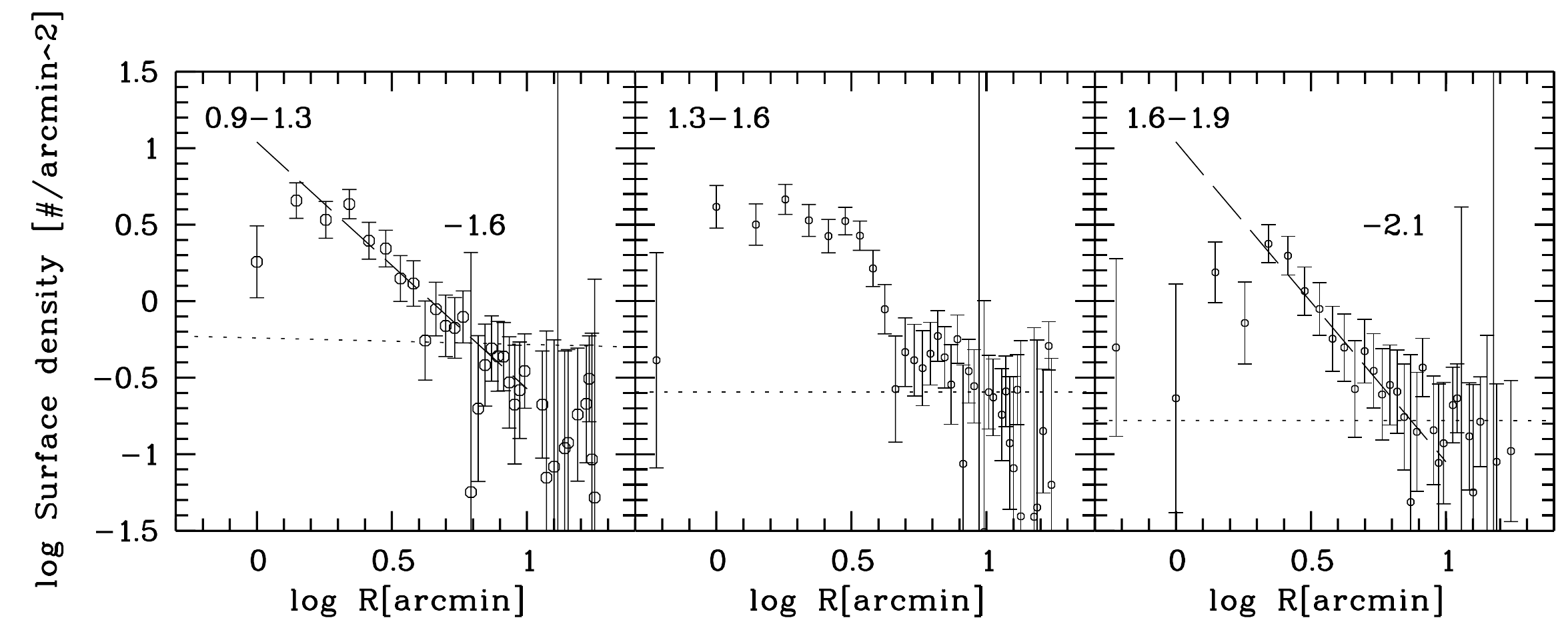}
\caption{Surface densities in double-logarithmic display for point sources with the background subtracted and for three different C-R color intervals:0.9-1.3 (left panel),
1.3-1.6 (middle panel),1.6-1.9(right panel). For the blue and the red samples, the power-law indices, valid for
the radial range between log(r)=0.4 and log(r)=0.9 are
indicated. There is no uniform power-law for the intermediate sample.  The dashed horizontal lines indicate the backgrounds. The vertical line is an excessively large uncertainty in one bin. It becomes
clear that the inflection point in Fig. \ref{fig:profile_all} is caused by the intermediate sample, which sharply falls off at a distance
of 4.5\arcmin. We identify this sample with the bulk
of intermediate-age clusters. }
\label{fig:profile_colors}
\end{center}
\end{figure*}

We now consider the radial distributions in different color intervals.
 This is done in Figs. \ref{fig:profile_density_linear} and  \ref{fig:profile_colors}.  Fig. \ref{fig:profile_density_linear} plots
the surface densities linearly and includes the color interval 0.8 $<$ C-R $<$ 1.1 where the clusters should be younger than
1 Gyr.  This population is well visible in Fig. \ref{fig:profile_density_linear}  at radii corresponding to the bulge region, and  less conspicous at larger
radii (due to the poor statistics we choose a larger bin in Fig.\ref{fig:profile_colors}).

In Fig. \ref{fig:profile_colors}, the left panel shows a blue sample (0.9 $<$ C-R $<$1.3), the middle panel an intermediate sample
(1.3 $<$ C-R $<$1.6), and the right panel (1.6$<$ C-R $<$1.9) the red sample. It becomes clear that the inflection point in Fig.\ref{fig:profile_all} is caused by the intermediate sample
whose density shows a rapid decline at 4.0\arcmin\  and remains more or less constant for larger radii. 
Since we are azimuthally averaging along structures which are azimuthally strongly inhomogeneous, this constancy
does not reflect a true radial constancy. 
Interestingly, the inflection point is not visible in the luminosity profile of NGC 1316.  
 Given all evidence, we identify this  
sample with the bulk of intermediate-age clusters.
 For the blue and the red sample, the indicated power-law indices refer to the
radial interval  0.4 $<$ log(r) $<$ 0.9. The red GC candidates show a somewhat steeper decline. They perhaps preserved their initially
more concentrated profile.

Fig.\ref{fig:xyplot_color} shows that the distribution of the intermediate sample at larger radii is concentrated towards the south,
roughly delineating Schweizer's L1-structure.



\section{Azimuthal distribution}

\begin{figure}[h!]
\begin{center}
\includegraphics[width=0.5\textwidth]{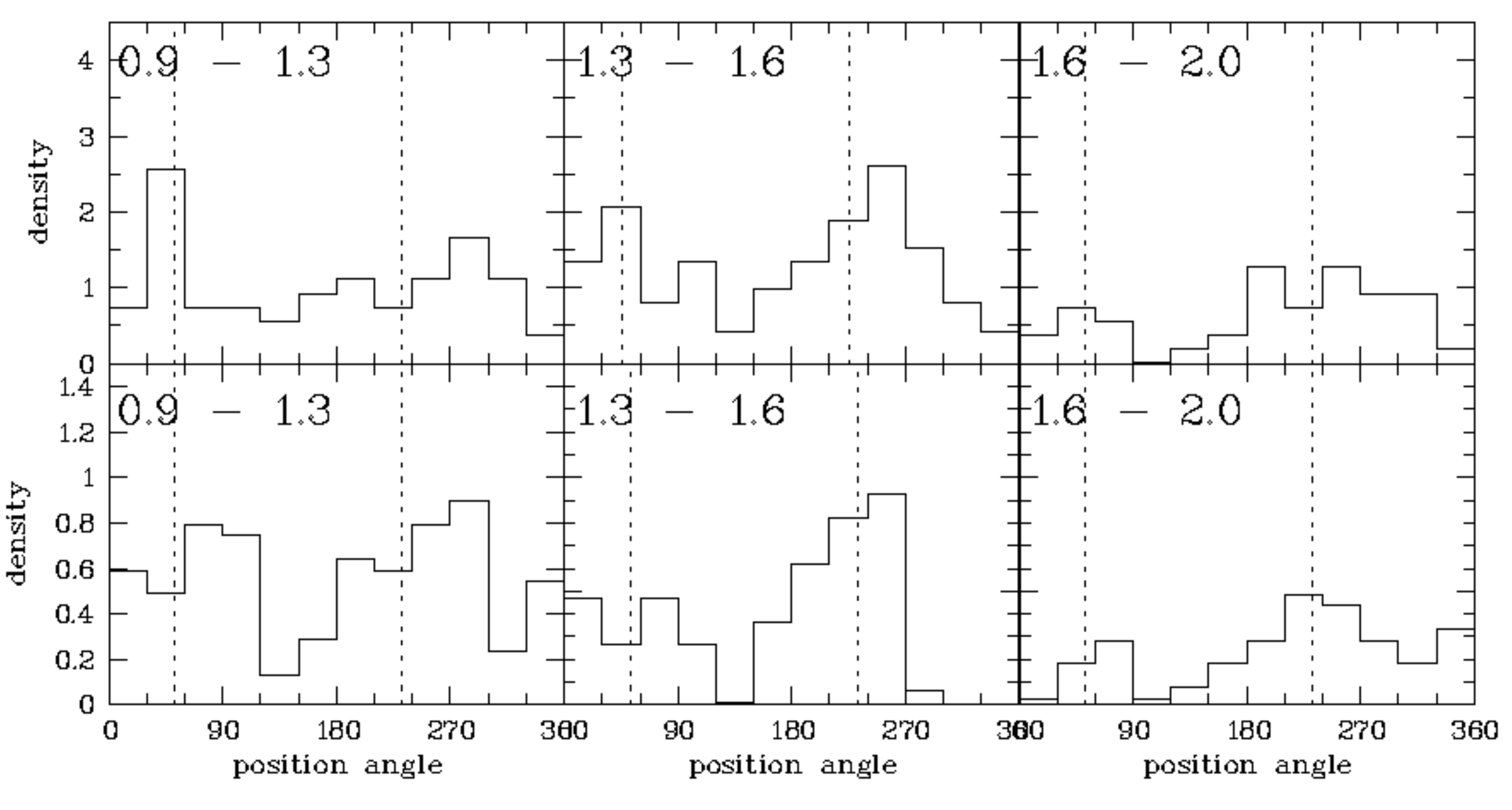}
\caption{The distribution of position angles in the radial ranges 2\arcmin\ - 4\arcmin\ (upper row) and 4\arcmin\- 10\arcmin\
(lower row) for three color intervals. The major axis is indicated by the vertical dotted lines. Sphericity is detected nowhere.}
\label{fig:azimut}
\end{center}
\end{figure}


Fig. \ref{fig:azimut} shows the azimuthal distribution for three color-subsamples of GC candidates for the radial
range 2\arcmin\ - 4\arcmin\ (inner sample) and  4\arcmin\ - 10\arcmin\ (outer sample). The bin width
is  30$^\circ$ and the counts have been performed in circular annuli. The abscissas are the position angles measured
counterclockwise with zero in the North. The ordinates are background-subtracted number densities per arcmin$^2$.
The first observation is that no subsample shows a  spherical distribution,  This stands in some contradiction to
\citet{gomez01} who found, using BRI colors,  for their blue sample a spherical distribution, while their red sample
followed the galaxy's ellipticity. We suspect that this is due to our more precise background subtraction and, because of the C filter inclusion, 
a more precise color subsampling. The highest degree of symmetry is found for the inner intermediate-color sample, once
more demonstrating the strong confinement  of this population to the bulge. The outer intermediate sample shows the
strong dominance of the southern region. Most of these clusters might be attributed to the L1 feature.
 The red sample, in which we expect old metal-rich clusters, also shows the bulge  symmetry.

\section{Luminosity function}

The luminosity function (LF) of the GCS has been derived to very faint magnitudes by HST \citep{goudfrooij01b}, although
only for the central parts. The main result is that the LF misses a well-defined turn-over magnitude (TOM) , which for ``normal'' 
elliptical galaxies is an excellent distance indicator \citep{villegas10,richtler03}.
This can be understood by the relative youth of the cluster system which contains still unevolved clusters, an effect
already noted by \citet{richtler03} in  galaxies with younger populations. 

Our data cannot compete in depth with the HST observations, but we can probe the outer parts of the GCS, where
our data are complete down to R=24 mag, in order to see whether a TOM is visible. With a distance modulus of
31.25 and a TOM of $\mathrm  M_V = -7.4$ we expect it at about V= 23.8 and with a mean V-R=0.5. At R=23.3, a TOM-brightness
 outside the main body of NGC 1316 should be clearly visible.

\begin{figure}[h!]
\begin{center}
\includegraphics[width=0.4\textwidth]{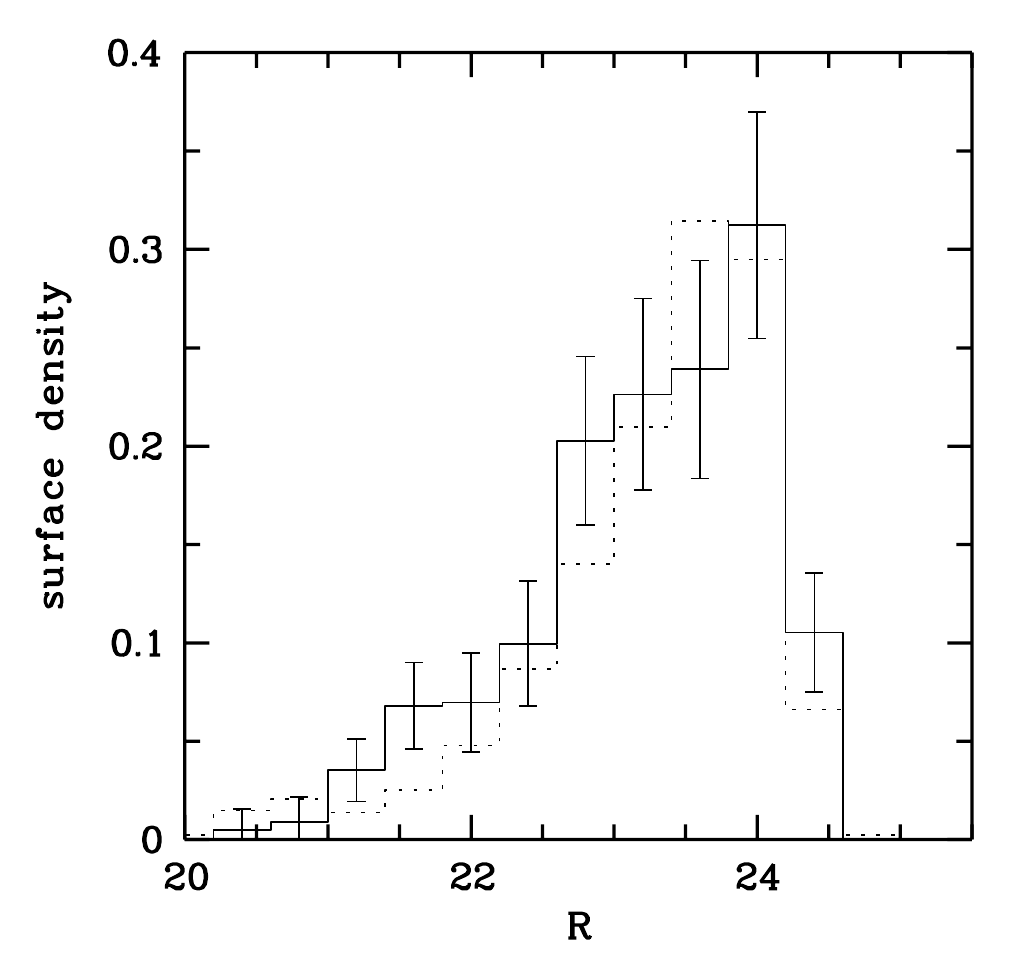}
\caption{The luminosity function in R inside a circular annulus between 4\arcmin\ and 9\arcmin\ radius. The surface
density is counted in numbers/arcmin**2.  The background counts have been performed outside 12\arcmin. We do not see the expected turn-over at R$\approx$ 23.3. One concludes
that also outside the main body of NGC 1316 younger clusters dominate the LF.}
\label{fig:LKF}
\end{center}
\end{figure}

As the radial interval we choose 4-9 arcmin\ in order to avoid the galaxy light, have reasonable number counts
and avoid the very outskirts. Fig. \ref{fig:LKF} shows the background-corrected LF in this radial range, where the background counts
refer to the complete field outside 12\arcmin. Surface densities are numbers/arcmin$^2$. The uncertainties are calculated as the square root of background 
corrected counts. The dotted histogram shows the background counts.

The expected turn-over at R$\approx$23.3 mag is not visible, resembling the HST results of \citet{goudfrooij04}  for the central part and for the
red clusters.
The conclusion is that  in this outer region the LF of  intermediate-age clusters contribute significantly.
The data of \citet{goudfrooij04} indicate a turn-over for their blue sample, but see our remarks in Sect.\ref{sec:discussion_color}
for counterarguments.

\section{Discussion}

\subsection{External and internal reddening}
\label{sec:reddening}
The foreground reddening toward NGC 1316 is low. The all-sky maps of
\citet{burstein82} and \citep{schlegel98} give E(B-V)=0 and
E(B-V)=0.02, respectively. This difference is the zero-point difference between the two
reddening laws.
According to Schlegel et al.,  the accuracy
of both values is comparable  below a reddening of E(B-V)=0.1. Therefore the
reddening in C-R is E(C-R)= 0.04 or probably smaller, comparable with
the uncertainty of the photometric calibration. We thus prefer not to correct
for foreground reddening, also given that no conclusion depends on such
a precision.  The reddening by dust within NGC 1316 outside a radius of 1\arcmin\ seems to be low as well.
We point to the fact that our reddest GC candidates have C-R$\approx$1.9, which the appropriate color for old, metal-rich
GCs.

\subsection{Theoretical Washington colors}
\begin{figure}[h!]
\begin{center}
\includegraphics[width=0.5\textwidth]{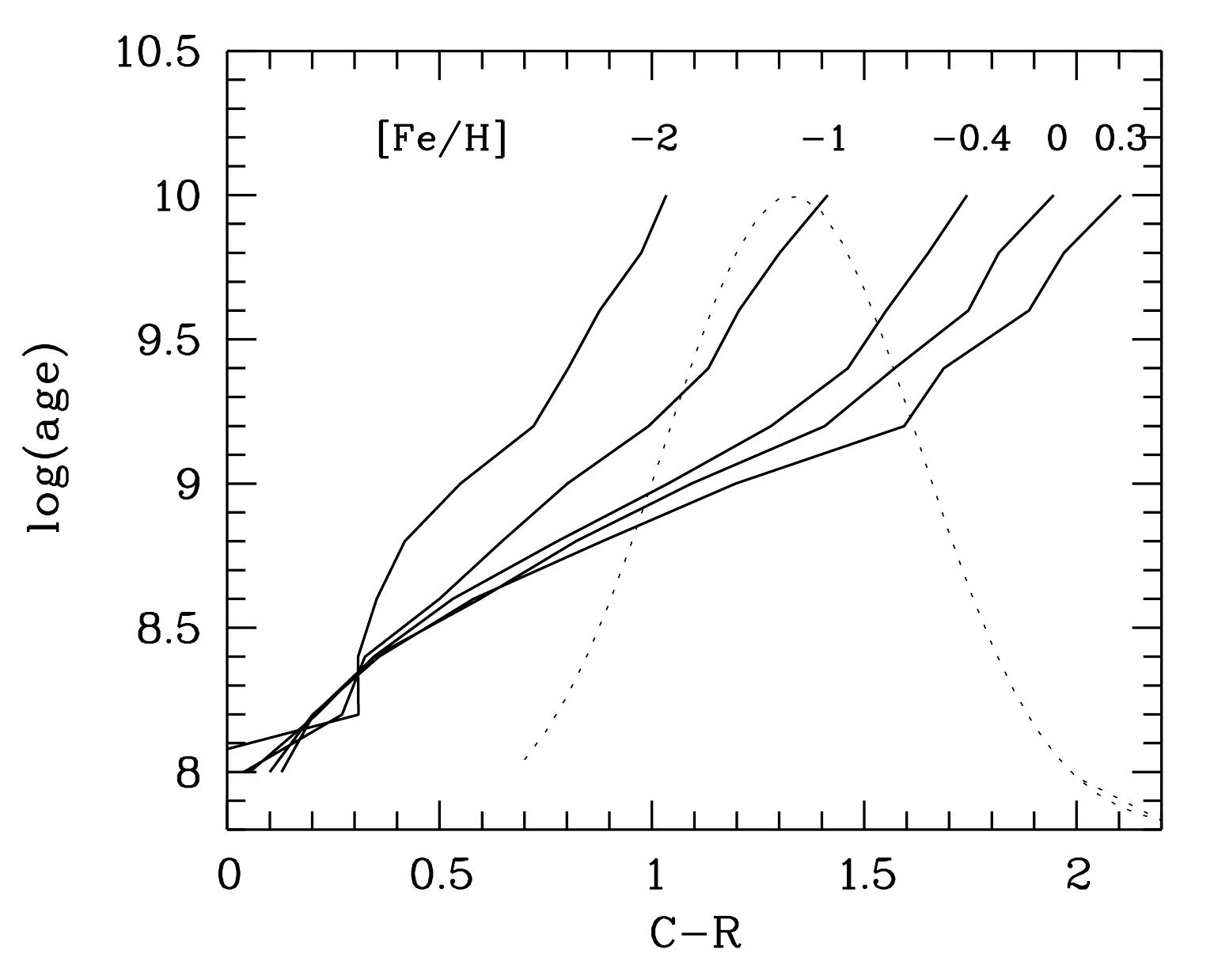}
\caption{This graph shows theoretical single stellar population models from \citet{marigo08} for five different metallicities. Overplotted (in arbitrary units) is a Gaussian representing the color distribution for GC candidates between 2 and 10 arcmin radius.   If the cluster sample within $1.3 < C-R < 1.6$ is dominated by objects with at least solar metallicity, many clusters are younger than 2 Gyr. }
\label{fig:washington}
\end{center}
\end{figure}

 We use  theoretical  isochrones  for relating cluster ages to Washington colors. 
Fig. \ref{fig:washington} compares the color distribution of GC candidates with theoretical models of single stellar populations from \citet{marigo08}, using their web-based tool (http://stev.oapd.inaf.it/cgi-bin/cmd).  Models are shown for five different metallicities. The
highest metallicity refers to the stellar population (e.g. \citealt{kuntschner00})
Overplotted (in arbitrary units) is a Gaussian representing the color distribution for GC candidates between 2\arcmin\ and 10\arcmin\ radius. 

\subsection{Color distribution and ages}
\label{sec:discussion_color}
What are the implications of  the color distribution in  Fig. \ref{fig:colordistribution}  for possible scenarios of the cluster formation history
of NGC 1316?  First we compare our C-R photometry with the HST-photometry in B,V,I  of  the cluster sample of \citet{goudfrooij01a}. 
These authors find two peaks at B-I=1.5 and B-I=1.8. Translating these values into Washington colors with the help of Fig. \ref{fig:comparison}, one estimates  for these peaks C-R=1.1 and C-R=1.4, respectively. This is indeed consistent with what is
seen in Fig. \ref{fig:colordistribution}. Already  \citet{goudfrooij01b} interpreted the peak at B-I=1.8 as the signature of intermediate-age
clusters with ages around 3 Gyr which  was backed up by the spectroscopic ages and metallicities derived for a few bright objects \citep{goudfrooij01a}.
From Fig.\ref{fig:washington} one would rather assign an age of somewhat less than 2 Gyr.
If the bluer peak  indicates old, metal-poor clusters of the pre-merger components, the metallicity
distribution of old clusters must be strongly biased to very metal-poor objects and thus be very different from that in ``normal'' globular cluster systems of ellipticals (and also spirals) which show
a  more or less universal peak at C-R = 1.35 \citep{bassino06b}.  

Since we can identify clusters with ages even younger than what would correspond to C-R=1.1, that means below 1 Gyr, one rather expects a mixture of old and
intermediate-age clusters of unknown proportions. The hypothesis that the blue peak does not represent an overabundance
of very metal-poor clusters, but marks a second burst of star formation with an age of about 1 Gyr, is interesting. A counterargument
could be that the luminosity function in a V,V-I diagram seems to show a turn-over at about V= 24.6 mag \citep{goudfrooij04}. However,
a number of points related to Fig.3 of Goudfrooij et al. create doubts whether this turn-over really corresponds to the universal turn-over magnitude in GCS. Firstly, \citet{goudfrooij04} adopted a distance of 22.9 Mpc which means a brightening  of 0.5 mag in absolute
magnitudes with respect to the supernova distance of 17.8 Mpc, we adopted. So the turn-over would be at $M_V = - 6.7$, i.e. almost 0.7 mag
fainter than the ''normal'' turn-over, while one would expect the turn-over to be brighter due to the bias to low metallicities. Secondly, the turn-over itself is not well defined, the highest number count being even fainter
than the formal turn-over from the fit, leaving the question open whether there is a decline at fainter magnitudes or not. Thirdly, their color interval which defines "old clusters" is $0.55 < V-I < 0.97$, a sample which
clearly includes intermediate-age clusters (here we also point to the superiority of the Washington system with respect to V-I and
refer the reader to Fig. 7 of   \citealt{dirsch03b} ).

Evidence from our photometry comes from the fact that the red peak is strongly constrained to the bulge (Figs. \ref{fig:profile_colors}
and \ref{fig:profile_density_linear}). The blue clusters, on the other hand, are abundant still in the outskirts of NGC 1316. If this outer cluster
population would consist exclusively of old clusters, then a turn-over magnitude should be visible in our luminosity function
which is not even corrected for incompleteness.

The absence of the striking red peak in the outer region shows that these stellar populations do not stem from the first star-burst.
The dynamical age of these regions is unknown, but plausibly younger than the inner star burst.

 The light of the inner bulge is redder than the peak color of the GCS by about  0.15 mag in C-R and shows only a small
radial transition  to bluer colors.   \footnote{Strictly speaking, this difference is not exactly valid,
because we measure the color of the stellar population by using the total projected light and calculate
a color for the GCs by averaging over magnitudes. The difference, however, is negligible.} 

A differential reddening between clusters and stellar population is not probable. The selection of cluster candidates between
1\arcmin\ and 3 \arcmin\ (236 objects), where the reddening should be strongest, still gives a peak color of C-R=1.38, while the galaxy has C-R=1.57. An explanation for this difference may be sought in the fact that the clusters are (more or less) single stellar populations of intermediate-age, while the total bulge light has a contribution from old 
 populations. We shall argue that this contribution is significant.

We find statistical evidence for clusters between $0.8 < C-R < 1.1$ which are neither old and metal-poor nor
can be related to a starburst 2 Gyr ago, but must be younger even if their metallicities are  lower than solar.
Plausibly,  these clusters should have a metallicity at least as high as the
intermediate-age stellar population, if they formed in NGC 1316.

However, there is still the possibility that infalling dwarf galaxies provided clusters with low metallicities


\subsection{Comparison with spectroscopic ages}
\citet{goudfrooij01b} provide spectroscopic ages and metallicities for three clusters, their objects No.103, No.114, and No. 210. There exist C-R measurements for 103 and 210 (114 is too close to the center). These clusters have solar abundance and a common
age (within the uncertainties) of 3 Gyr. The C-R colors for 103 and 210 are 1.48 and 1.51, respectively. A comparison with Fig.\ref{fig:washington} shows that these colors are a bit too blue for full consistency with the model colors for 3 Gyr and solar
abundance.  A   shift to the red of 0.1 mag in C-R would be necessary to achieve perfect agreement, but given that various models
with their respective uncertainties and the uncertainty of the absolute photometric calibration are involved, one can consider
the agreement to be satisfactory. In any case, these two clusters do not seem to be  reddened.

\subsection{Subpopulations}
\label{sec:subpop}
A bimodal color distribution is not exclusively a feature of elliptical galaxies, as the example of the Milky Way shows.
As a working hypothesis, we assume that the color histogram of old  GCs of NGC 1316 also has a bimodal appearance. Moreover, we
assume that  clusters redder than the galaxy color belong to the old, metal-rich subpopulation.   
\citet{bassino06b} compiles values for the blue and the red peak of early-type galaxies in the Washington system. Adopting these
values, we fix the Gaussian of the red peak with a maximum at C-R=1.75 and a dispersion of 0.15.   We cannot distinguish
old, metal-poor clusters from younger, metal-rich clusters that populate the same color range, but may assume that their
color distribution follows the almost universal distribution known from other galaxies, for which \citet{bassino06b}
quotes C-R=1.32  for the peak color  and 0.15 mag for the dispersion.

In order to get an impression of what the pre-merger GCSs could look like, we  fit the corresponding  Gaussian to the color bins redder than C-R=1.6, assuming that 
all clusters redder than the galaxy light are old. The amount of old, metal-poor clusters is unknown, but we assume two  cases: equal to
and twice the metal-rich population. The subtraction of these two Gaussians from the Gaussian describing the entire 
sample down to R=24 mag in Sect. \ref{sec:color} 
reveals the GC population which has been formed in the merger and and perhaps in following star-forming events.    

\begin{figure}[h!]
\begin{center}
\includegraphics[width=0.5\textwidth]{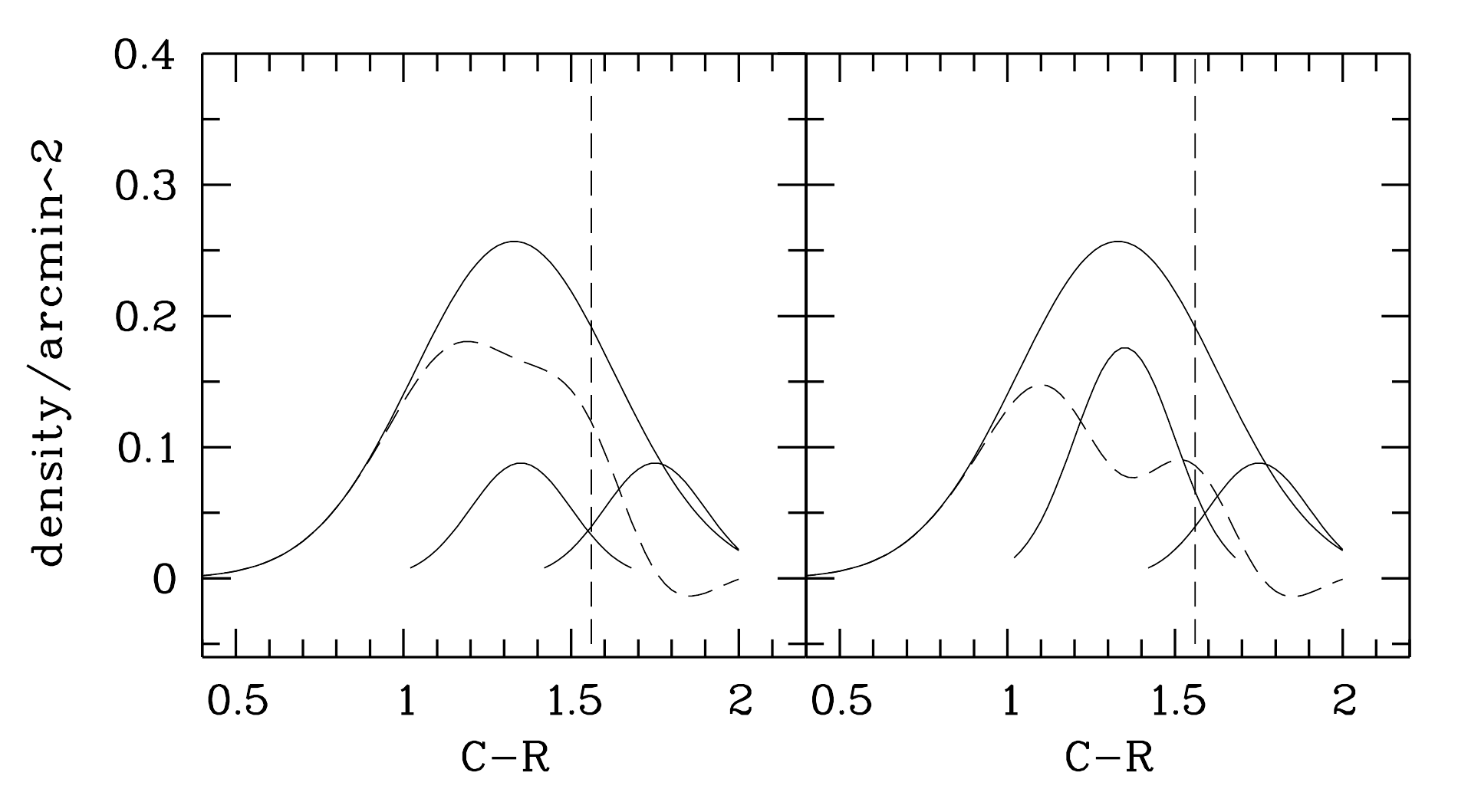}
\caption{In both panels, the broad Gaussian  (solid line) represents the color distribution of the entire GCS. The two narrower Gaussians
represent the assumed bimodality of old metal-poor and metal-rich clusters. The dashed line results from subtracting the bimodal
color distribution.  In the left panel (case 1), intermediate-age clusters dominate, in the right panel (case 2), old clusters dominate.
Case 2 is less consistent with the properties of the GC population.  }
\label{fig:subpop}
\end{center}
\end{figure}

Fig. \ref{fig:subpop} shows these two cases. In the left panel (case 1) the dashed line results from the subtraction of the assumed
bimodal color distribution of old clusters. In this case, intermediate-age clusters are dominating the color interval 1.3-1.6. In the
the right panel (case 2), old clusters dominate which contradicts the behaviour of clusters in this color range. Therefore, case 1
may be closer to the true situation.

\subsection{A star formation history}
\label{sec:scenario}

The star formation history of NGC 1316 was probably complex.  A merger event is not a simple infall, but is preceded by close encounters which can trigger
starbursts, in time separated by the orbital period.   See, for example, \citet{teyssier10} and \citet{dimatteo08} for modern simulations of mergers and starbursts.
Teyssier et al. resolve the multi-phase ISM and find that the main trigger for starbursts is not a large-scale gas flow, but a fragmentation  of gas into cold and dense
clumps, favoring star cluster formation. A realistic scenario would probably involve many individual starbursts which perhaps can be identified in future high S/N spectra
of  globular clusters in NGC 1316.  Our intention in simplifying the star formation history is to show that  in order to bring photometric properties and dynamical determined
mass-to-light ratios into agreement, the properties of the pre-merger population rule out an elliptical galaxy. Moreover, a contribution of  a population significantly
younger than 2 Gyr is needed.

 We therefore ask whether it is possible to find a simple model, inspired by the color distribution of GCs, which is consistent
with the color and the mass-to-light ratio of the bulge stellar population.
We adopt three populations: a
pre-merger population (pop.1) with
an a-priori unknown age, and two populations representing starbursts with ages 2 Gyr (pop. 2) and 0.8 Gyr (pop. 3). 
Since the luminosity weighted metallicity of the bulge is at least solar \citep{kuntschner00}, we adopt  solar metallicity for each of them. 
The color of NGC 1316 then results from a composite of these populations to which we assign
the integrated color C-R=1.55 (see  Fig.\ref{fig:colorprofile}).

 Mass-to-light ratios for the inner bulge have been quoted by several authors in various photometric bands. To transform an M/L-value from
 one band to another, is in the case of NGC 1316 not completely trivial, since  published aperture photometries \citep{prugniel98} measure 
 a color which is reddened by the dust within about 1\arcmin\ radius. We therefore adopt theoretical colors (including Washington colors) from
 the Padova models  (http://stev.oapd.inaf.it/cgi-bin/cmd; \citealt{marigo08}). 
  The relation between M/L-ratios (denoted by $\Gamma$) in filters i and j is
 $$\frac{\Gamma_i}{\Gamma_j} = \frac{L_j}{L_i} \frac{L_{i,\odot}}{L_{j,\odot}}  $$
 C-R=1.55 corresponds to a good approximation to a single stellar population with solar metallicity and an age of 2.5 Gyr. For that population we find B-V=0.81, B-R=1.34,
 and R-I=0.56. Furthermore, we adopt the solar absolute magnitudes of \citet{binney98} (p.53) and thus can transform M/L-values into the R-band. 
 
   \citet{schweizer81} gives $\mathrm M/L_V = 1.3 \pm 0.2$ for a distance
 of 32.7 Mpc, corresponding to  $\mathrm M/L_R = 2.1$ for our distance (M/L is inversely proportional to the distance).
 \citet{shaya96} give $M/L_V \approx 2.2 \pm 0.2$, corresponding to $M/L_R \approx 1.9$ (distance 16.9 Mpc).  
 \citet{arnaboldi98}, by using dynamics of planetary nebulae, quote $\mathrm M/L_B \approx 7.7$ inside a radius of
200\arcsec\ (distance 16.9 Mpc).
 However, they find  lower values for smaller radii, down to   $\mathrm M/L_B \approx 4.3$ at a radius
of 45\arcsec\ which corresponds to $\mathrm M/L_R \approx 3.1$.
This early result, based on only a few planetary nebulae, is superseded by the recent work of \citet{mcneil12}, who quote $\mathrm M/L_B = 2.8$ (distance 21.5 Mpc)
for the stellar population, which transforms into $\mathrm M/L_R \approx 2.6$. 
Due to the large database of their dynamical model, this value certainly has a high weight. However, it is based on spherical modeling and the authors caution that
it may change with more realistic models. 
    \citet{nowak08}, modeling the central kinematics with a black hole, have found for the central region a value in the $K_s$-band between
0.7 and 0.8 (distance 18.6 Mpc), corresponding to a value between 2.8 and 3.2 in the R-band (setting $M_Ks/M_{Ks,\odot}$ = $M_K/M_{K,\odot}$ and adopting 
R-K=2.60 for the stellar population).
Future work will probably lower the uncertainties.  For the moment, $M/L_R \approx 2.5$ seems to be a good value to adopt.
 
 Perhaps the best empirical determinations of M/L-values of
stellar populations of early-type galaxies come from the SAURON-group \citep{cappellari06}.  The highest values, they
quote in the I-band inside one effective radius, are around 5 (their "Jeans"-values). Considering that a dark halo can contribute
up to 8\% of the total mass, values around 4 for metal-rich, old populations seem reasonable. The Padova models give $M/L_I$=4.1 for a 12 Gyr old 
population of solar metallicity (with $M_{I,\odot}$=+4.10), assuming a log-normal IMF \citep{chabrier01}, reproducing the empirical values quite well.

Adopting their M/L-ratios also for younger populations, we now can assign colors and M/L-values to our model populations
and estimate under simplifying assumptions the properties of the pre-merger population.



Table \ref{tab:pops} lists the models, ages, metallicities, Washington colors, and R-band mass-to-light-ratios (columns 1-5).
We assume two different pre-merger populations with ages of 12 Gyr (pop.1a) and 7 Gyr (pop.1b).
 The equation to be solved is then

$$\mathrm  (C-R)_{integrated} = 1.55 = -2.5 \cdot log \left (\frac{(L1+L2+L3)_C}{(L1+L2+L3)_R} \right ) $$

where L1, L2, L3 are the respective luminosities in C and R. If we use the R-luminosity of the bulge  as unit and
assume a certain mass proportion of pop.2 and pop.3, one can solve for L2 and L3. Since in case 1 of Fig. \ref{fig:subpop} the young clusters
are of comparable number, we assume pop.2 and pop.3 to contribute with
equal stellar mass. L2 and L3 (R-band) are given in column 6. Column 7 gives the resulting stellar masses for the bulge
(inside a radius of 240\arcsec\ ) using the bulge luminosity of Table \ref{tab:photmodel}. These values are  upper limits since
in reality one must use the deprojected bulge luminosities, but the global symmetry of NGC 1316 is not known.

Having these values, one can
calculate the resulting M/L of the composite population.


\begin{table}[h!]
\caption{Properties of  three adopted single  stellar populations and their composite}
\begin{center}
\resizebox{9cm}{!}{
\begin{tabular}{ccccccc}
\hline
Pop. &  Z & age[Gyr] & C-R  & $M/L_R$ &$\mathrm  L_R$& $\mathrm M[M_\odot] $ \\
 (1)   &  (2) &   (3)  & (4) & (5) & (6) & (7) \\
\hline 
1a &  0.019& 12 &1.99 & 5.2  & 0.60  & $\mathrm  3.18 \times 10^{11}$ \\
2 &  0.019 &2 &1.49 & 1.1    & 0.13 &  $\mathrm 1.55 \times 10^{10}$\\
3 & 0.019 & 0.8 & 0.96 & 0.57  & 0.27 & $\mathrm 1.53 \times10^{10}$\\
 \hline
 composite &   &  &1.55 &3.4& 1 & $\mathrm 3.48 \times 10^{11}$  \\
\hline
\hline
1b & 0.019 &  7 &  1.84   & 3.4 & 0.67 & $\mathrm 2.36 \times 10^{11}$\\
2 &  0.019 &2 &1.49 & 1.1    & 0.11 &  $\mathrm 1.22 \times 10^{10}$\\
3 & 0.019 & 0.8 & 0.96 & 0.57  & 0.22 &  $\mathrm 1.26 \times 10^{10}$\\
\hline
composite  & & & 1.55 & 2.55 &1 & $\mathrm 2.60 \times 10^{11}$\\
\hline
\hline
\end{tabular}
}
\end{center}
\label{tab:pops}
\end{table}

 What one learns from this exercise is   that a pre-merger 
 population of 12 Gyr with solar metallicity, representative of an elliptical galaxy, does not agree well with a composite $M/L_R$=2.5.
In contrast, a 7 Gyr old population fits well,  which one would expect if the pre-merger was one or two  spiral galaxies with an already
existing mix of old to intermediate-age populations.

Moreover, a population significantly younger than 2 Gyr is necessary in case of $M/L_R$=2.5. If only the 2 Gyr population would exist, it would be so dominant ($\mathrm L2_R=0.85$ 
in the case of a 12 Gyr pre-merger population) that $\mathrm M/L_R=1.7$. This value would even be lower in the case of a younger pre-merger population.

These considerations are viable only for the bulge. In the outer parts, the galaxy gets bluer, but we have no knowledge of
the metallicities or ages. However, one might suspect that younger ages dominate over decreasing metallicity, since the 
Southern L1-structure is traced by GCs of probable intermediate age.
 

 We thus have a total stellar bulge mass of about $2.6\times10^{11}$ $M_\odot$ of which only
10\% have
been produced in both starbursts. These 10\% are, however, responsible for
the majority of bright (!) GCs, emphasizing once more that the efficiency of
GC formation in phases of high star formation rates is greatly enhanced \citep{whitmore95,larsen00,degrijs03a,degrijs03b,kravtsov05}.



 One has to account for a gas mass of $2.6\times10^{10}$ $M_\odot$ which has been transformed into
stars. Since we must assume a strong galactic wind with heavy mass loss during the burst phases, the total gas mass needed was probably higher. A gas mass of this order is easily provided by one or two
 spiral galaxies (e.g. \citealt{mcgaugh97}).




\section{Conclusions}
We present wide-field photometry (36 $\times$ 36 arcmin$^2$) in Washington C and Kron-Cousins R  around the
merger remnant NGC 1316 (Fornax A) in order to investigate the globular cluster system  of NGC 1316 on a larger scale
than has been done before. The data consist of MOSAIC images obtained with the 4-m Blanco telescope at
CTIO. The Washington system is particularly interesting due to its good metallicity resolution in GCSs containing
old clusters and for the comparison with the well-established bimodal color distribution of elliptical galaxies.

Our main findings are:\\
The GC candidates are well confined to the region of the optical extension of NGC 1316. Outside a radius of
12\arcmin\ (corresponding to 62 kpc), a cluster population is statistically not detectable.\\
The entire system down to  R = 24 mag shows  a broad 
Gaussian-like distribution with a peak at C-R = 1.37, somewhat bluer than the color of the inner bulge. A selection of brighter GC candidates  shows  a bimodal color distribution,  best visible for bulge objects, with peaks
at C-R=1.1 and C-R=1.4.   We find a small population of GC candidates  bluer
than C-R=1.0, which is the limit for old metal-poor clusters. Clusters bluer than C-R=0.8 are statistically not
detectable, but casual confirmations through radial velocities show that there are sporadically clusters
as blue as C-R=0.4. The brighter population of cluster candidates therefore consists of intermediate-age clusters,
while older clusters of the pre-merger galaxy are progressively mixed in at fainter magnitudes.     

Assuming that younger cluster have at least the metallicity of the bulge population, one can assign ages, using
theoretical Washington isochrones and integrated colors. The peak at C-R=1.4 then corresponds to an age of 
about 1.8 Gyr, the peak at C-R=1.1 to an age of 0.8 Gyr, and the bluest colors to less than 0.5 Gyr.
The peaks plausibly stem from epochs of very high star formation rates, connected to one or several  merger events.
The bluest clusters could indicate ongoing star formation younger than 0.8 Gyr. Alternatively, they could be
metal-poorer clusters of somewhat older age from infalling dwarf galaxies or could be formed in one of the star bursts from metal-poor gas of   one of the pre-merger components. 

The intermediate color interval $1.3 < C-R < 1.6$ shows a radial surface density profile very different from the other color regimes. Here we
find the bulk of intermediate-age clusters. These objects are strongly confined to the inner bulge, showing an inflection at about 4\arcmin.
For radii larger than 4\arcmin, their distribution is azimuthally  inhomogeneous with a
pronounced concentration in the area of Schweizer's (1980) L1-structure.
 The blue clusters ($0.9 < C-R < 1.3$)
fall off without showing an inflection  with a power-law exponent of -1.6. The red clusters ($1.6 < C-R < 2$) follow a somewhat steeper
power-law with an exponent of -2.1.

Outside the bulge, the luminosity function of cluster candidates does not show the turn-over expected for an old
cluster system, indicating that also at larger radii, younger clusters contribute significantly.

 
 Guided  by the color distribution of GCs, we present a simple model, which as the main ingredient adopts two burst of star formation
 with ages 2 Gyr and 0.8 Gyr added to an older population of solar metallicity. In order to reproduce the color of the bulge of NGC 1316, and the stellar M/L-value from dynamical estimates, this older population has an age of about 7 Gyr as a single stellar population. In reality, this population is expected to be a mix of populations and thus the
 age indicates spiral galaxies as merger components rather than an old elliptical galaxy.  The existence of a population significantly younger than 2 Gyr
 is necessary in order to avoid the dominance of this population which would result in too low M/L-values.
 
  The stellar masses of the younger populations account for only 10\% of the total stellar mass, while their luminosity contributes  30\%  of the total luminosity.
  
 
 As an appendix, we present our image  and add morphological remarks. 
 The wide-field morphology of NGC 1316 does not reveal new features with respect to the photographic morphological study
of \citet{schweizer80} with the exception of a faint arc which morphologically could be a continuation of the prominent L1-feature.
We also present a color map of the inner regions  showing  details, which to our knowledge have not been mentioned in the literature, among them a curved feature which may be the relic of a spiral arm.


The color gradient is very shallow out to 6\arcmin\
pointing to a well mixed stellar population.
We construct a new spherical model of the R-brightness profile and give its main characteristics. 
 \begin{acknowledgements}
We thank an anonymous referee for valuable remarks and constructive criticism.
TR acknowledges financial support from the Chilean Center for Astrophysics,
FONDAP Nr. 15010003,  from FONDECYT project Nr. 1100620, and
from the BASAL Centro de Astrofisica y Tecnologias
Afines (CATA) PFB-06/2007. He also thanks the Arryabhatta Institute for Observational Sciences, Nainital, for warm hospitality 
and financial support. LPB gratefully acknowledges support
by grants from Consejo Nacional de Investigaciones Cient\'ificas
y T\'ecnicas and Universidad Nacional de La Plata (Argentina).
We thank Richard Lane for a careful reading of a draft version.

\end{acknowledgements}

\appendix
\section{Morphology of NGC 1316}
The morphology of NGC 1316 has been discussed several times (e.g. \citealt{schweizer80,mackie98}), 
 but to our knowledge, a wide-field deep CCD image of NGC 1316 does not exist in the literature. Moreover, 
 the color map of the inner region exhibits details, which are worth to be shown. 

\begin{figure*}[t]
\begin{center}
\includegraphics[width=0.7\textwidth]{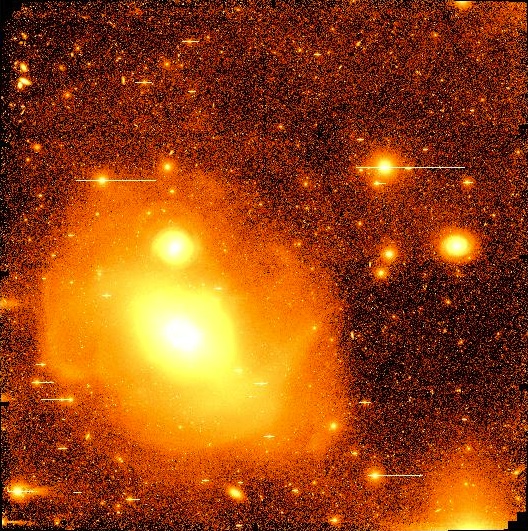}
\caption{Our R-image of NGC 1316. The size is 36\arcmin $\times$ 36\arcmin. North is up, East to the left.
One notes the relatively well defined outer border. The diameter along the major axis is 24.5 \arcmin, along
the minor axis 21.5\arcmin. The intensity levels are displayed logarithmically. Almost all faint structures
are already visible on the photographic image of \citet{schweizer80} except the very faint arc which appears
as a continuation of the plume reaching out to the NW. The faint larger-scale structures in the upper part
are probably flat field structures. There is scattered light into the camera at the lower right edge, coming
from the V=7.7 mag star HD 20914.
The border of the globular cluster system corresponds well to the visible luminous body.} 
\label{fig:picture}
\end{center}
\end{figure*}




\subsection{Global morphology}

A deep wide-field image of NGC 1316 reveals a wealth of faint structures. This paper has the 
emphasis on GCs, but it is appropriate to make a few remarks on the global morphology. 
\citet{schweizer80} gave a thorough morphological description of the NGC 1316 complex on
a scale of about 1 degree. For orientation and labels of the various features, we refer the reader
to Fig.2 of \citet{schweizer80}. 


It is interesting to see that our deep CCD-image (Fig. \ref{fig:picture}) is
not much superior to photographic Schmidt  plates regarding the detection of faint areal structures.
 In Fig. \ref{fig:picture} almost all features labeled by Schweizer are visible. There
are only two structures which we do not have in common. One is the very faint arc
to the SE (Schweizer's Fig.9) which is outside our field. The other is an
extremely faint curved arc (resembling a ``spiral arm"), which
appears as a continuation of the L1 structure 
reaching  out to the NW. Regarding this feature, one may
 think of the ``rows" in grand design spiral galaxies where parts of spiral arms are straight lines rather than
curved. This has not been discussed much in the literature. \citet{chernin98} gives a gas dynamical interpretation,
but encounters may also play a role.

For example, the structure of the grand design spiral M51 has been analyzed by \citet{dobbs10} through hydrodynamical models
who identified tidal influences (in the case of M51 the interaction with NGC 5195) as a main driver for shaping spiral
structure. In analogy, NGC 1316 could have interacted with its companion NGC 1317 or with a later merged galaxy.  


Intuitively, the region outside the bulge
 gives the impression of 
 an inclined disk rotating clockwise. Practically all loops are consistent with this sense of rotation
except the outermost loop in the SW. However, a bulge component kinematically dominates outside 35\arcsec\
\citep{bedregal06}.

Very noticable is the sharp border in the South. This may be another hint that we are looking onto a (thick?)
disk since 3-dimensional structures in projection should not be so sharply confined.

 The distribution of GCs is well confined to the luminous area seen in Fig.\ref{fig:picture}.
We emphasize the morphological  similarity of NGC 1316 with NGC 474/470. \citet{sikkema06} presents V-I photometry of its
GCS and only finds a broad blue peak, similar to our entire sample. 

 \subsection{The surface brightness profile}
 Due to the inner saturated region, we can measure the surface brightness profile of NGC 1316 in the R-band  on our MOSAIC image only from
 about 50\arcsec outwards. In order to enable the comparison with published aperture photometry, we use the ellipse-task in IRAF to measure (almost)
  circular  ``isophotes" and try to find an analytic spherical model for the surface brightness. We integrate this radial model and calculate aperture photometric values
 which we compare with the compilation of aperture photometries of \citet{prugniel98}. For the model, we chose a  ''beta-model'' which turns out
 to work surprisingly well. The exponent of -1  enables  deprojection and integrated mass to be written analytically (which here would result in only approximate values,
 since the system is not spherical) (e.g. see \citealt{schuberth10,richtler11}).

 \begin{equation}
\label{eq:light1}
\mu(R)=-2.5\log \left(a_1 \left (1+\left( \frac{R}{R_c} \right)^2 \right)^{\alpha}\right ) 
\end{equation}
with $a_1$\,=\,3.902$\times$10$^{-7}$, $R_c$\,=\,8\arcsec,
 $\alpha$\,=\,$-1.0$.  
To transform
 this surface brightness into  $L_\odot/pc^2$, one has to apply a factor $2.56\times 10^{10}$ to the argument of the logarithm (where we used $M_{R,\odot} = 4.45$). 
 The foreground extinction in the R-band is 0.056 mag (\citealt{schlegel98}) which we neglect.
 
\begin{figure}
\begin{center}
\includegraphics[width=0.35\textwidth]{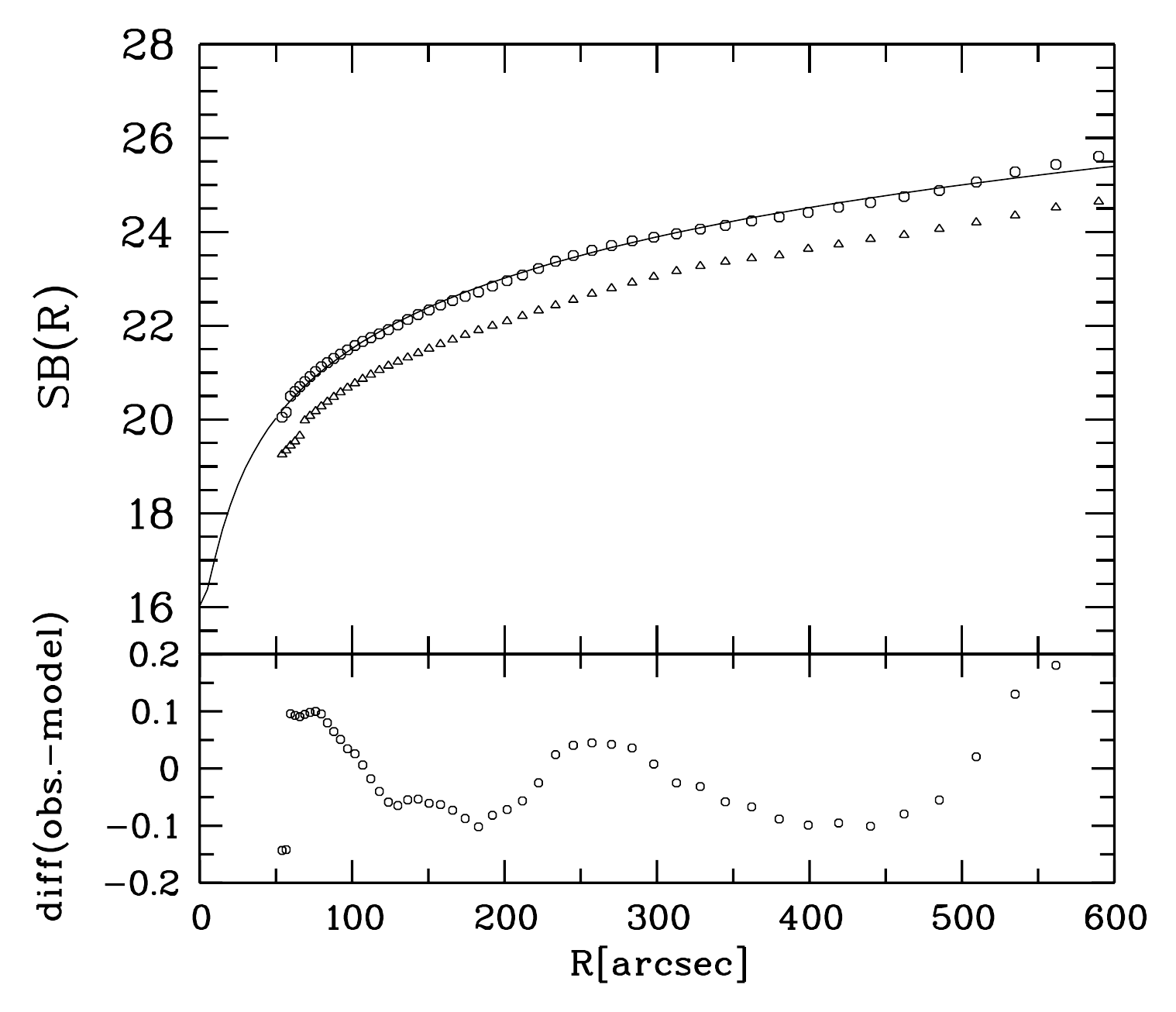}
\caption{Upper panel: Our surface-brightness profile in the R-band. The solid line is the beta-model. Open circles denote the surface brightness with {\it spherical}
isophotes. Open triangles denote the surface brightness with elliptical isophotes (ellipticity 0.3, position angle  50$^\circ$). This profile has been shifted by -0.5 mag
for better visibility.  
 Lower panel: residuals in mag between our measurements (upper panel) and
the photometric model in the sense: observations$-$model.}
\label{fig:light_profile}
\end{center}
\end{figure}

\begin{figure}
\begin{center}
\includegraphics[width=0.35\textwidth]{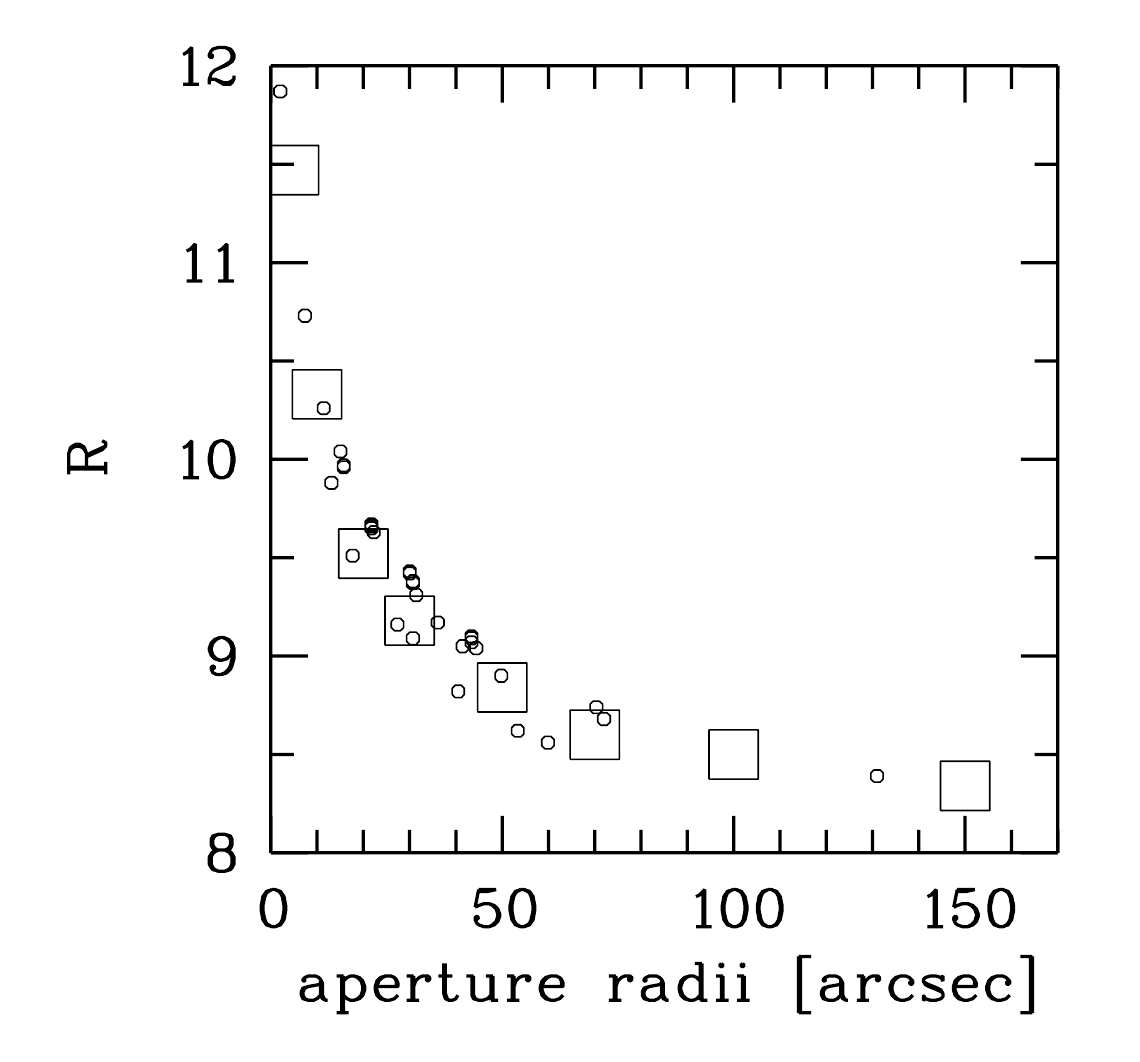}
\caption{ This graph shows that the model of our surface-brightness profile reproduces  published R-band aperture photometries excellently. x-axis shows  aperture radii, the y-axis the apparent R magnitude. It is astonishing that the extrapolation of the beta-model into the inner region works so well.
  Open circles  are from the compilation of \citet{prugniel98}, squares are simulated aperture photometries of our beta-model.
   }
\label{fig:apertures}
\end{center}
\end{figure}

 Fig.\ref{fig:light_profile} shows in its upper panel the beta-model (solid line), the spherical model (open circles), and measurements
 of elliptical isophotes (ellipticity 0.3, position angle 50$^\circ$). These measurements have been shifted by 0.5 mag to enhance
 the visibility. In the spherical model, the bulge does not emerge as a separate entity, because the sphericity introduces some
 smoothing. Also in the elliptical isophotes, it is hardly visible.  We compare our model with published aperture photometry \citep{prugniel98} by integrating within 
 given apertures.  The result is shown in Fig. \ref{fig:apertures} for eight apertures and is found to be very satisfactory.

 \begin{table}[h!]
\caption{Characteristic values of the spherical photometric model}
\begin{center}
\resizebox{5cm}{!}{
\begin{tabular}{lll}
\hline
 Radius  &   $M_R$  &  $L_R/L_\odot  $  \\
\hline
 240\arcsec\  (bulge) & -23.07 & $\mathrm 1.02 \times 10^{11}$ \\
600\arcsec\  & -23.33 & $\mathrm 1.3 \times 10^{11}$ \\
\hline
\end{tabular}
}
\end{center}
\label{tab:photmodel}
\end{table}

 Table \ref{tab:photmodel} lists the absolute   R-magnitudes and corresponding luminosities for the bulge and the total extension of NGC 1316.
 The effective radius is 69.5\arcsec, corresponding to almost exactly 6 kpc.
 
 
\subsection{Color image}
 \begin{figure*}
\begin{center}
\includegraphics[width=0.7\textwidth]{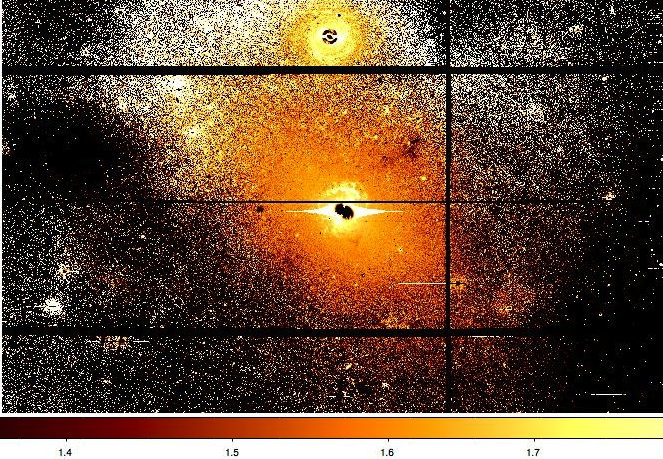}
\caption{A color image of NGC 1316. See text for details.   }
\label{fig:color}
\end{center}
\end{figure*}

The color image is restricted in size because of the strong flatfield features present on the western part of the C-image.

Fig.\ref{fig:color} displays the color in magnitudes. Its size is 23.3\arcmin $\times$ 14.5\arcmin. After subtracting the sky, we divided the C-image by the
R-image and converted the resulting image to a  magnitude scale.
We selected from the GC photometry 10 of the brightest non-saturated objects at small radii and applied their calibrated
magnitudes 
to the galaxy light to make sure that the galaxy colors are measured differentially to the GC colors.
The resulting scatter is 0.03 mag in C-R. The dynamical range of Fig.\ref{fig:color} covers the color
interval 1.3 - 1.9, which has been found the most satisfactory. Dark is blue and bright is red. The magnitude scale itself is
displayed logarithmically. 
Due to the uncertain flat-field of the C-image at low intensity levels,
 the colors beyond about 8\arcmin\ are not reliable. Particularly, we are reluctant to claim the reality of the outer
red structures in the North, but mention that on Schweizer's plate (his Fig.1) the Northern extension of
NGC 1316 appears more diffuse than the Southern border.

The extended line emission region of \citet{mackie98} is visible as the bright spot, presumably dust, in the South
just above the CCD-gap, which unfortunately covers part of the structure.
The southern HII-region, detected by \citet{schweizer80} appears as a dark spot not well visible in this display. It
is an extremely interesting object, apparently showing globular cluster formation under quite isolated conditions.
We intend to devote an own contribution to this object.

\subsection{The color profile}
\begin{figure}[h]
\begin{center}
\includegraphics[width=0.3\textwidth]{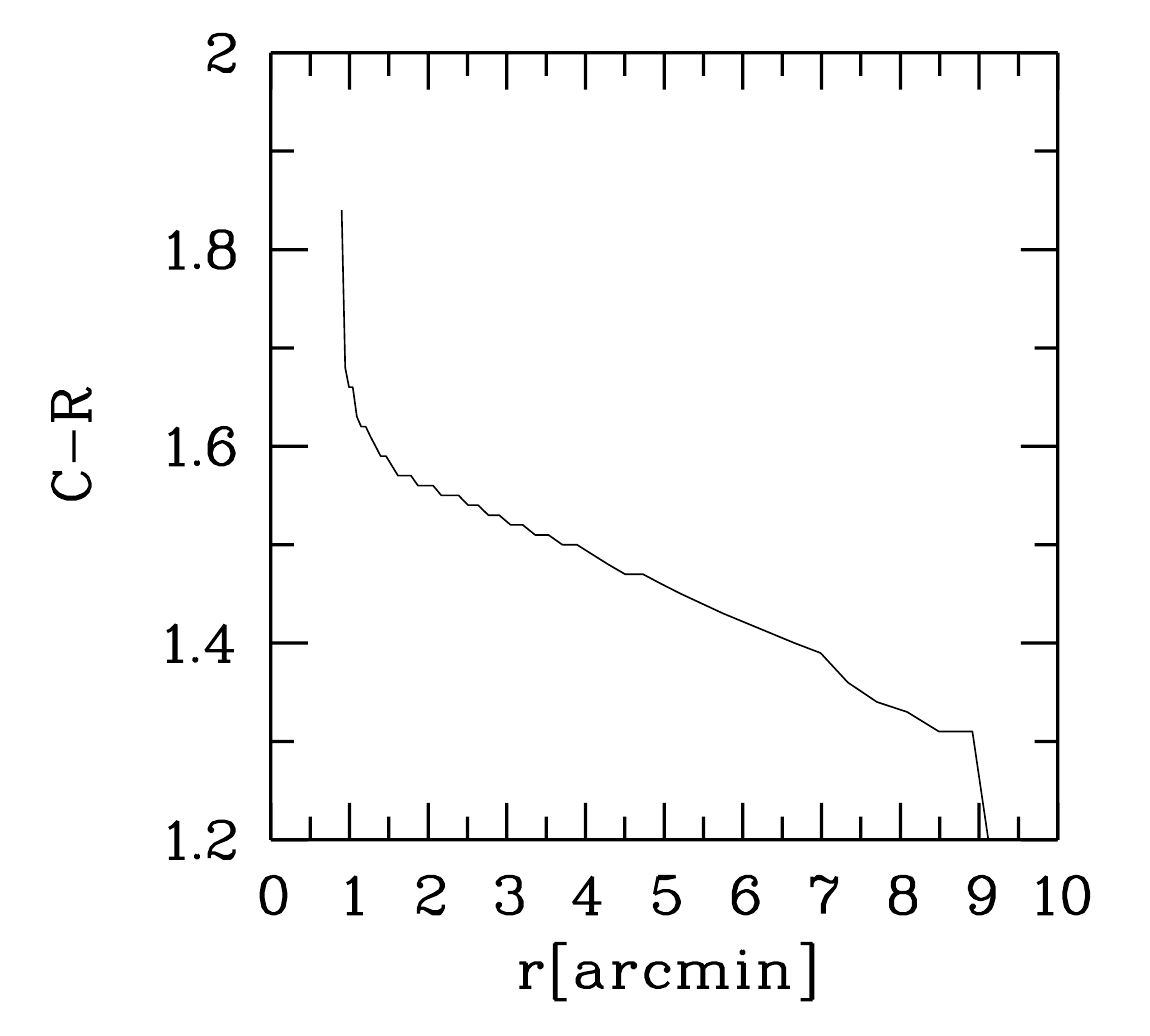}
\caption{The radial color profile.  The color has been measured along the major axis of NGC 1316  averaging  along elliptical isophotes   with a position angle of 50$^\circ$.
 }
\label{fig:colorprofile}
\end{center}
\end{figure}

The color map does not exhibit a clear elliptical appearance. In spite of that we adopt the ellipticity of the optical isophotes and 
evaluate the color by averaging  along ''color-isophotes'' with a fixed ellipticity of 0.3 and a position angle of 50$^\circ$, corresponding to the major
axis of NGC 1316. Fig.\ref{fig:colorprofile} shows the radial dependence of color. The inner jump to very red colors is due to the central dust structures. The gradient has been noted already by \citet{schweizer80} who  attributed it to a declining abundance in the outer
parts.   The question, whether age or abundance is the dominant factor, must be left for future investigations  
  of the outer clusters.

\subsection{Color image of the central parts}

\begin{figure*}[h]
\begin{center}
\includegraphics[width=0.7\textwidth]{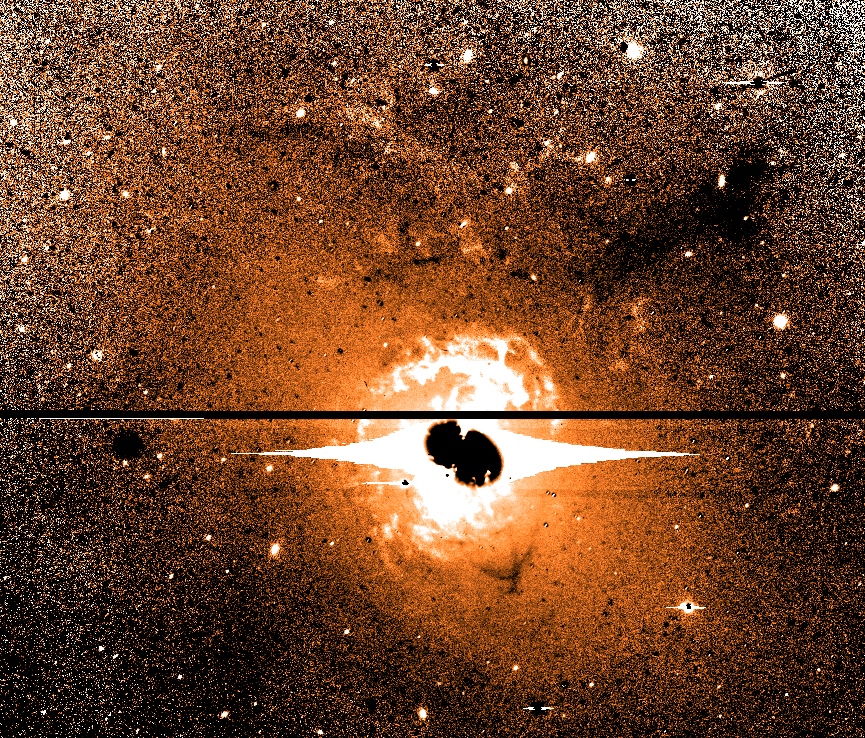}
\caption{The inner part of NGC 1316 showing the C-R color map. The size is 7.8\arcmin $\times$ 6.5\arcmin. Red is
bright, blue is dark. The dynamical range is 1.5 $<$ C-R $< $1.7, so also tiny color differences become visible. See text for details. }
\label{fig:color_center}
\end{center}
\end{figure*}

The wealth of structure and features  which is visible in the inner color map, deserves a special description. The basis is Fig.\ref{fig:color_center}.
This map displays C-R in the range from 1.5 to 1.7. White is red, dark is blue. The reddest colors are caused by dust, while bluer
colors are caused by bluer stellar populations and/or  perhaps by strong emission of the OII-3727 line which falls into the C-filter. 
The inner dust features best known from HST images appear white with a color about C-R = 1.9. They
fit well  to non-stellar PAH emission \citep{temi05,kaneda07}. Note that the overall symmetry of the dust distribution differs from the
spheroid. It shows a slightly elliptical distribution with its major axis at a PA of about -30$^\circ$. 

Note the many tiny dust patches in the north which are absent in the south. Remarkable is  the slightly curved brighter feature, beginning at about
2.5\arcmin ~to the North, the PA first being 70$^\circ$ and 90$^\circ$ on its eastern part. The overall impression is that of a relic of a spiral arm, which may be a direct clue to the nature of the original galaxy. At its southern border, it is accompanied by
a more blueish color, reminding of the fact that dust usually follows trailing spiral arms.

One of the most conspicuous features is Schweizer's ''plume'', pointing towards the NW. Neither Schweizer nor  \citet{mackie98} 
detected line emission. No atomic or molecular gas has been detected \citep{horellou01}. Dust also seems to be absent. Its total extent in the NW-direction is almost 2\arcmin, corresponding
to 10 kpc. One finds the bluest color at the "head" in some blue spots which have C-R=1.1. Towards the SE-direction, the color
becomes somewhat redder, typically 1.4. Since we observe in projection, this does not mean that the color of the associated
stellar population is really varying, but only that its density varies. The true color might be quite blue and homogenous.
Unless this population has a low metallicity (an infalling dwarf galaxy in tidal dissolution cannot be excluded), it must be younger than 1 Gyr. 
\citet{graham87} made the interesting suggestion that the plume might represent young stars whose formation was triggered by a nuclear jet. 
A correspondence  might be seen in a group of blue stars in CenA that has it probably origin in star formation triggered by the X-ray jet \citep{graham02}.
However, its  major axis does  point neither to the nucleus nor to the center of the western inner radio lobe (which is not necessarily a counterargument
after roughly a crossing time). The   alternative (also mentioned by \citealt{graham87}) is an infalling  dwarf galaxy in dissolution,
which appears more probable.

Striking is also
the psi-shaped irregular structure 1\arcmin\ to the south, which is bluer than its environment by  0.05 mag which to our knowledge
has not yet been mentioned in the literature. Whether the color is due to a bluer population or due to line emission, must be cleared
up by spectroscopy.

\bibliographystyle{aa}
\bibliography{N1316}

\end{document}